 \documentclass[final,authoryear,5p,times,twocolumn]{elsarticle}
 
\usepackage{graphicx}

\usepackage{amssymb,amsmath}

\journal{Planetary and Space Science}

\begin{document}

\begin{frontmatter}



\title{Dust abundance and grain size in galaxy halos}


\author[label1]{Hiroyuki Hirashita}
\ead{hirashita@asiaa.sinica.edu.tw}
\author[label1,label2]{Chih-Yu Lin}
\address[label1]{Institute of Astronomy and Astrophysics,
Academia Sinica, P.O. Box 23-141, Taipei 10617, Taiwan}
\address[label2]{Department of Physics, National Chung Hsing University,
145 Xingda Rd., South Dist., Taichung 40227, Taiwan}

\begin{abstract}
We investigate the abundance and properties (especially, grain size)
of dust in galaxy halos using available observational data in the
literature. There are two major sets of data. One is (i) the
reddening curves at redshifts $z\sim 1$ and 2 derived for
Mg \textsc{ii} absorbers, which are assumed to trace the medium
in galaxy halos. The
other is (ii) the cosmic extinction up to $z\sim 2$ mainly traced by distant
background quasars. For (i), the observed reddening curves favor
a grain radius of $a\sim 0.03~\mu$m for silicate, while
graphite is not supported because of its strong
2175 \AA\ bump. Using amorphous carbon improves
the fit to the reddening curves compared with graphite if the grain radius is
$a\lesssim 0.03~\mu$m.
For (ii), the cosmic extinction requires $\eta\gtrsim 10^{-2}$
($\eta$ is the ratio of the halo dust mass to the stellar mass; the observationally
suggested value is $\eta\sim 10^{-3}$) for silicate
if $a\sim 0.03~\mu$m as suggested by the reddening curve
constraint.
Thus, for silicate, we do not find any grain radius that satisfies both (i) and
(ii) unless the halo dust abundance is much larger
than suggested by the observations. For amorphous carbon, in contrast,
a wide range of grain radius ($a\sim 0.01$--0.3~$\mu$m) is accepted
by the cosmic extinction;
thus, we find that a grain radius range of $a\sim 0.01$--0.03 $\mu$m
is supported by combining (i) and (ii).
We also discuss the origin of dust in galaxy halos,
focusing on the importance of grain size in the physical mechanism of
dust supply to galaxy halos.
\end{abstract}


\end{frontmatter}


\section{Introduction}
\label{sec:intro}

The interstellar dust is known to occupy around 1\% of
the total mass of the interstellar medium (ISM) in
the Milky Way, which means that around half of
the metals\footnote{We refer to
elements heavier than helium (such as C, N, O, etc.) as metals,
following the convention in astronomy.}
are in the solid (dust) phase \citep[e.g.][]{Evans:1994aa,Jenkins:2009aa}.
Dust forms and evolves mainly in the
ISM of galaxies through various processes \citep{Asano:2013aa}.
Since a galaxy is
not a closed system, the ISM interacts with the circum-galactic
medium (CGM) and the intergalactic medium (IGM) through outflow
driven by supernovae (SNe) and active galactic nuclei (AGNs)
\citep[e.g.,][]{Veilleux:2005aa},
and through inflow driven by cooling and/or gravity
\citep[e.g.,][]{Keres:2005aa}. The outflow
could also transport the interstellar dust to the CGM and IGM
\citep{Zu:2011aa,McKinnon:2016aa,Hou:2017aa,Aoyama:2018aa}.
Radiation pressure
from newly formed stars in a galaxy could also drive the interstellar
dust outward, supplying the dust to the CGM and IGM
\citep{Ferrara:1991aa,Bianchi:2005aa,Bekki:2015ab}.
Therefore, the cosmic dust has a wide spatial distribution not limited
to the ISM but extended over the cosmological
volume \citep[see also][]{Aguirre:1999aa}.

The dust content in galaxy halos (or the CGM; hereafter, we simply
use words ``galaxy halo'' to indicate the circum-galactic environment
up to a radius of $\sim 100$--200 kpc)
gives us a clue to the transport of dust from the ISM to the IGM, since
galaxy halos are the interface between the ISM and the IGM.
\citet[][hereafter M10]{Menard:2010aa} detected
reddening\footnote{In this paper, we define the reddening as the
difference between the dust extinctions (usually expressed in units
of magnitude) at two wavelengths. Since
dust extinction tends to be larger at shorter wavelengths,
the color usually becomes redder than the intrinsic one after dust
extinction.}
in galaxy halos using the cross-correlation between
the galaxy position and the reddening of background quasi-stellar objects
(QSOs) for
a large sample of galaxies taken by the Sloan Digital Sky Survey
(SDSS; \citealt{York:2000aa}). The median redshift of their
sample is $z\sim 0.3$ ($z$ denotes the redshift).
They detected reddening up to
a radius of several Mpc from the galaxy center.
\citet{Peek:2015aa} applied basically the same method to nearby
galaxies ($z\sim 0.05$), and found a similar radial profile of
reddening to the one found in M10.
{\citet{Masaki:2012aa} confirmed the observationally suggested
large extent of dust distribution in galaxy halos by comparing their
analytic halo model with M10's data.}

{The existence of dust in galaxy halos is important in the following
aspects. Dust contained in galaxy halos
on a line of sight causes dust extinction on a cosmological distance scale.
This could lead to a systematic reddening effect on distant QSOs and SNe.
Dust in galaxy halos is also of fundamental importance in the total dust budget
in galaxies and in the Universe, since M10 estimate that
the dust mass in a galaxy halo is on average
comparable to that in a galaxy disc.}
Moreover, as \citet{Inoue:2003ab,Inoue:2004aa} argued,
dust in the IGM could affect the thermal state of the IGM through
photoelectric heating. They constrained the grain size and
the dust-to-gas ratio in the IGM using the observed thermal
history of the IGM, although those two quantities are
degenerate in such a way that small grains require
smaller abundance of dust. It is also important to note that
dust extinction strongly
depends on grain size (or grain size distribution)
and dust-to-gas ratio
\citep{Inoue:2004aa,Inoue:2010aa}. Thus, among
dust grain properties, we focus on grain size and dust-to-gas ratio
(or dust abundance) in this paper.

To clarify the dust properties in galaxy halos, it would be
observationally convenient to
target specific objects that represent the CGM.
\citet[][hereafter MF12]{Menard:2012aa} argued that Mg\,\textsc{ii}
absorbers trace the gas in galaxy halos, based on the impact
parameters. The estimated dust-to-gas ratio for Mg\,\textsc{ii}
absorbers is also consistent with that expected for $L^*$
galaxies, which indicates that Mg\,\textsc{ii} absorbers
are associated with $\sim L^*$ galaxies
(i.e., galaxies with the characteristic optical luminosity
in the luminosity function) \citep{Menard:2009aa}.
Deep imaging observations also support this association
\citep{Steidel:1997aa,Zibetti:2007aa}. MF12 pointed out that,
if Mg\,\textsc{ii} absorbers are associated with outflow
originating from metal-enriched galaxies,
their dust-to-gas ratios comparable to the Milky Way
dust-to-gas ratio ($\sim 1/100$) can be naturally explained.
Although we should keep in mind that the origin of Mg\,\textsc{ii}
absorbers has still been debated, there are
some observations that showed a link between Mg\,\textsc{ii}
absorbers and outflows driven by stellar feedback (energy input from
stars) \citep{Bond:2001aa,Tremonti:2007aa,Bouche:2007aa}.

\citet{York:2006aa} studied about 800 Mg \textsc{ii} absorbers at
$z\simeq 1.0$--1.9. By
comparing sub-samples with various absorption strength, they derived
the extinction curves for those absorbers, finding that they are described
well by the Small Magellanic Cloud (SMC) extinction curve.
MF12 analyzed the reddening of background QSOs and derived
reddening curves (reddening as a function of rest-frame wavelengths)
for Mg\,\textsc{ii} absorbers at $z\sim 1$ and 2.
They found that the reddening curves are fitted well with the
SMC extinction curve. For nearby
($z\sim 0.05$) galaxies,
\citet{Peek:2015aa} derived a similar reddening
curve to the one obtained above, except for the strong excess
in the $u$-band (i.e., at the shortest among their sampled wavelengths).
The strongly rising trend toward short wavelengths ($\lambda\lesssim 0.2~\mu$m,
where $\lambda$ is the rest-frame wavelength)
in the SMC extinction
curve, which is consistent with the reddening curves of Mg \textsc{ii}
absorbers, indicates the existence of
grains with radii smaller than $\lambda /(2\pi )
\sim 0.03~\mu$m \citep{Bohren:1983aa}. In other words, the
reddening curves could be used to obtain the information on
grain size in galaxy halos.

In this paper, we aim at putting a constraint on the dust grain size
and the dust abundance in galaxy halos based on the currently
available data in the literature. The obtained constraints on
those quantities could be used to clarify the origin and evolution
of dust in galaxy halos. Although our analysis is limited by the
current uncertainty in the observational data, we make a first
attempt to draw
a useful conclusion about the grain size as we see later.
We also discuss some theoretical predictions in the literature on the
dust properties in galaxy halos, and comment on
the limitation of the current knowledge on dust in galaxy halos.

The paper is organized as follows. We formulate the model of
reddening curves and dust extinction in Section~\ref{sec:model}.
We show the results and compare them with observational data
in Section~\ref{sec:result}. We further discuss
the model predictions in terms of theoretical predictions in
the literature, and clarify the limitation of our knowledge on
halo dust in Section \ref{sec:discussion}.
Finally we conclude in Section \ref{sec:conclusion}.
Throughout this paper, we adopt $H_0=70$ km s$^{-1}$ Mpc$^{-1}$,
$\Omega_\mathrm{M}=0.3$, and $\Omega_\Lambda =0.7$ for the
cosmological parameters.

\section{Model}\label{sec:model}

We construct a model that describes the reddening caused by
dust in galaxy halos. We mainly treat the wavelength dependence of
reddening, which we refer to as the \textit{reddening curve}, and
the extinction over the cosmic distance, which we refer to as the
\textit{cosmic extinction}. We formulate
those quantities in what follows.

\subsection{Reddening}\label{subsec:model_reddening}

The wavelength dependence of extinction can be used to constrain
dust properties, especially grain size.
Although it is difficult to measure the extinction in a single band,
the color excess or reddening, which is a relative flux change at two
wavelengths due to dust extinction, is rather easily derived.
The reddening is defined as the difference in the extinctions at
$\lambda_1$ and $\lambda_2$ ($\lambda_1>\lambda_2$).

The currently available data for
dust in galaxy halos are not precise enough to constrain the
detailed grain size distribution.
{The functional form of the grain size distribution
and the maximum and minimum of grain radius are degenerate
(i.e., it is usually difficult to determine all those properties
uniquely), especially when the reddening is given with a sparse
sampling of wavelengths.}
Thus, we assume that all the dust grains have a single
radius, $a$, for simplicity.
This simple assumption is useful
to obtain the typical
radius of dust grains that dominate the opacity in galaxy halos.
{We additionally discuss the effect of grain size distribution in
Section \ref{subsec:sizedist}.}

To evaluate the extinction, we need the mass extinction coefficient
(the cross-section for extinction per dust mass),
$\kappa_\mathrm{ext}(\lambda )$ \citep{Hildebrand:1983aa}:
\begin{eqnarray}
\kappa_\mathrm{ext}(\lambda )=\frac{3Q_\mathrm{ext}(\lambda,\, a)}{4as},
\label{eq:kappa}
\end{eqnarray}
where $Q_\mathrm{ext}(\lambda,\, a)$ is the extinction cross-section
per geometric cross-section, and the grains are assumed to be spherical
with grain radius $a$ and material density $s$.
{We calculate $Q_\mathrm{ext}$ using the Mie theory with
grain properties in the literature (Section \ref{subsec:param}).}
The extinction at wavelength $\lambda$, $A_\lambda$, is estimated as
\begin{eqnarray}
A_\lambda =2.5(\log\mathrm{e})\,\kappa_\mathrm{ext}(\lambda )\mu m_\mathrm{H}
N_\mathrm{H}\mathcal{D} ,\label{eq:A_lambda}
\end{eqnarray}
where $\mu (=1.4)$ is the gas mass per hydrogen, $m_\mathrm{H}$ is the
mass of hydrogen atom, $N_\mathrm{H}$ is the
column density of hydrogen nuclei, and $\mathcal{D}$ is the dust-to-gas ratio.

{Since the reddening $A_{\lambda_1}-A_{\lambda_2}$ is proportional to
$N_\mathrm{H}\mathcal{D}$,\footnote{{Extinction curves are often expressed
with normalization to the value at a certain wavelength $\lambda_0$:
$A_\lambda /A_{\lambda_0}$ \citep[e.g.][]{Pei:1992aa}. However, the
absolute value of extinction $A_{\lambda_0}$ is necessary and is
unknown for the observed extinction curve in MF12.
Reddening ($A_\lambda -A_{\lambda_0}$) is easier to derive from the QSO colors.}}
we need to adopt objects whose column density and
dust-to-gas ratio are well constrained.}
Following MF12, we use Mg \textsc{ii} absorbers as tracers of
dust in galaxy halos (see the Introduction).
According to MF12, the typical column density of an Mg \textsc{ii} absorber
is $N_\mathrm{H}\sim 10^{19.5}$~cm$^{-2}$, and the dust-to-gas ratio is
60--80 per cent of the Milky Way value if we use
$A_V/N_\mathrm{H}$ for the indicator of dust-to-gas ratio.
Assuming the typical dust-to-gas
ratio of the Milky Way to be 0.01 (or slightly less) \citep[e.g.,][]{Pei:1992aa},
we adopt $\mathcal{D}\sim 0.006$ for Mg \textsc{ii} absorbers.
\citet{Peek:2015aa} derived a reddening curve for galaxy halos at
$z\sim 0.05$. Their reddening curve is similar to the one obtained by
MF12, except for the strong excess of extinction in the $u$-band
(i.e., at the shortest wavelength among their bands).
As we see later, the column density, which is poorly known for
galaxy halos, is also important. Because of the well estimated
column density for Mg \textsc{ii} absorbers, we adopt the data in MF12.

\subsection{Extinction over the cosmic distance (cosmic extinction)}
\label{subsec:model_ext}

Dust enrichment in galactic halos has not been fully understood yet.
Thus, as a starting point, we simply scale the halo dust mass
with the stellar mass in the galaxy. This may be a reasonable starting
point because the dust is most probably supplied to the halo as a result
of stellar feedback (MF12; \citealt{Hou:2017aa}).
{\citet{Masaki:2012aa} also adopted this assumption in their
model that reproduced a consistent dust distribution to M10's
observation data.} From that assumption,
the dust mass in the halo, $M_\mathrm{d,halo}$ can be expressed as
\begin{eqnarray}
M_\mathrm{d,halo}=\eta M_\mathrm{star},\label{eq:eta}
\end{eqnarray}
where $M_\mathrm{star}$ is the stellar mass and $\eta$ is the
mass ratio of the dust in the halo to stars.
\citet{Peek:2015aa} showed, using a sample at $z\sim 0.05$, that
$M_\mathrm{d,halo}\propto M_\mathrm{star}^{0.23}$. Thus, in general,
$\eta$ depends on $M_\mathrm{star}$.

The statistics of the stellar mass per comoving volume in the Universe
is described by the stellar mass function
(probability distribution function of galaxies with stellar mass $M_\mathrm{star}$),
$\Phi (M_\mathrm{star})$.
We adopt the stellar mass function at $z < 3$ from \citet{Tomczak:2014aa}.
We estimate the comoving halo dust density, $\rho_\mathrm{d,halo}$, as
\begin{eqnarray}
\rho_\mathrm{d,halo} & = & \int M_\mathrm{d,halo}\Phi (M_\mathrm{star})\,
\mathrm{d}M_\mathrm{star},\nonumber\\
& = & \langle\eta\rangle\rho_\mathrm{star},\label{eq:eta_av}
\end{eqnarray}
where the integration range for the stellar mass is the same
($10^9\,M_\odot <M_\mathrm{star}<10^{13}\, M_\odot$) as in
\citet{Tomczak:2014aa}, $\langle\eta\rangle$ is the average
of $\eta$ for the corresponding stellar mass range, and
$\rho_\mathrm{star}$ is the comoving stellar
mass density (\ref{app:comoving}).
We adopt the following fitting for $\rho_\mathrm{star}$ as a function
of redshift $z$ as \citep{Tomczak:2014aa}
\begin{eqnarray}
\log\rho_\mathrm{star}~[M_\odot~\mathrm{Mpc}^{-3}]=-0.33(1+z)+8.75.
\end{eqnarray}

If we assume that the cosmic extinction up to redshift $z$
at wavelength $\lambda_\mathrm{obs}$ in the observer's frame
(the optical depth of the cosmic extinction is denoted as
$\tau_\mathrm{d,cosmo}(z,\,\lambda_\mathrm{obs})$) is dominated
by the dust in galaxy halos, we can estimate the dust optical depth
per comoving length as
$K_\mathrm{ext}=\sigma_\mathrm{g}n_\mathrm{g}\tau_\mathrm{d,halo}(\lambda )$,
where $\sigma_\mathrm{g}$ is the mean geometrical cross-section of galaxy halo,
$n_\mathrm{g}$ is the comoving number density of galaxies,
and $\tau_\mathrm{d,halo}(\lambda )$ is the mean extinction optical depth of galaxy halo
as a function of rest-frame wavelength $\lambda$.
At the same time, using the mass extinction coefficient,
$\tau_\mathrm{d,halo}(\lambda )$ [$\lambda$ is the rest-frame wavelength, which is
related to $\lambda_\mathrm{obs}$ as $\lambda =\lambda_\mathrm{obs}/(1+z)$]
is estimated as
$\tau_\mathrm{d,halo}(\lambda )=\kappa_\mathrm{ext}(\lambda )\rho_\mathrm{d} l$,
where $\rho_\mathrm{d}$ is the mean dust mass density
in galaxy halos and $l$ is the path length in a halo.
Thus, we obtain
$K_\mathrm{ext}=\sigma_\mathrm{g}n_\mathrm{g}\kappa_\mathrm{ext}(\lambda )\rho_\mathrm{d}l$.
Since $\sigma_\mathrm{g}l$ is
the mean volume of
galaxy halo, $\rho_\mathrm{d}\sigma_\mathrm{g}l$ is reduced to
the total dust mass per halo, denoted as $\langle M_\mathrm{d,halo}\rangle$
(\ref{app:comoving}).
After all, using equation (\ref{eq:rho_dhalo}), $K_\mathrm{ext}$ is expressed as
\begin{eqnarray}
K_\mathrm{ext}(z,\lambda )=
\kappa_\mathrm{ext}(\lambda )\langle M_\mathrm{d,halo}\rangle n_\mathrm{g}
=\kappa_\mathrm{ext}(\lambda )\rho_\mathrm{d,halo}(z).
\end{eqnarray}
{Note that, since we are interested in cosmological-volume properties
sampled by random lines of sight, only the mean quantities within a halo is
relevant here. Possible clumpiness of dust distribution in halos does not affect
our results. This is why the resulting opacity, $K_\mathrm{ext}$, depends
only on the mean dust mass density, $\rho_\mathrm{d,halo}$.}

Finally, the extinction optical depth up to redshift $z$ at wavelength
$\lambda_\mathrm{obs}$ in the observer's frame,
$\tau_\mathrm{d,cosmo}(z,\,\lambda_\mathrm{obs})$, is estimated as (M10)
\begin{eqnarray}
\tau_\mathrm{d,cosmo}(z,\,\lambda_\mathrm{obs})=\int_0^zK_\mathrm{ext}\left( z',\,
\frac{\lambda_\mathrm{obs}}{1+z'}\right)\frac{c(1+z')^2}{H(z')}\,\mathrm{d}z',
\end{eqnarray}
where $c$ is the light speed, and $H(z)$ is the Hubble parameter at $z$.
For the flat Universe,
\begin{eqnarray}
H(z)=H_0\left[\Omega_\mathrm{M}(1+z)^3+\Omega_\Lambda\right]^{1/2}.
\end{eqnarray}

\subsection{Choice of parameter values}\label{subsec:param}

The dust abundance in a galaxy halo is regulated by $\langle\eta\rangle$
(equation \ref{eq:eta_av}) in our model.
Hereafter, we simply denote $\langle\eta\rangle$ as $\eta$ for the brevity
of notation. M10 show that the dust mass
in a galaxy halo is on average $\sim 5\times 10^7~M_\odot$.
They also argue that the effective stellar luminosity of their sample
is $0.45 L^*$.
If we assume that the stellar mass of a galaxy with stellar luminosity $L^*$ is
the characteristic mass of the stellar mass function ($M^* = 10^{11.05}~M_\odot$)
derived by \citet{Tomczak:2014aa}, the mean stellar mass
in M10's sample is estimated as $\sim 0.45M^*=5\times 10^{10}~M_\odot$.
By taking the ratio of the typical dust mass in a halo
to the typical stellar mass, we obtain $\eta\sim 10^{-3}$ for
M10's sample. \citet{Peek:2015aa} derived a similar halo dust mass
for sub-$L^*$ galaxies. Thus, we adopt $\eta =10^{-3}$ as a fiducial value.

We consider silicate and graphite as representative grain materials
that reproduce the extinction curves in nearby galaxies
\citep{Draine:1984aa,Weingartner:2001aa}. In addition, we adopt
amorphous carbon \citep{Zubko:1996aa}, which could be required to
reproduce the extinction curves whose 2175 \AA\ bump
is not prominent \citep{Nozawa:2015aa,Hou:2016aa}.
We adopt $s=3.5$, 2.24 \citep{Weingartner:2001aa} and
1.81 g cm$^{-3}$ \citep{Zubko:2004aa} for the material densities of
silicate, graphite, and amorphous carbon, respectively.
{The extinction coefficient, $Q_\mathrm{ext}(\lambda ,\, a)$, in
equation (\ref{eq:kappa}) is calculated using the Mie theory
\citep{Bohren:1983aa} for spherical grains with the optical properties available in
the above references.}
\citet{Jones:2013aa} adopted a different line of models for
carbon dust species and also adopted compound (or coagulated) species.
However, the present observational data for
dust in galaxy halos are not rich enough to distinguish the detailed
dust properties; thus, we simply examine the difference between
graphite and amorphous carbon as two ``representative'' carbonaceous
species.

\section{Results}\label{sec:result}

We compare our calculated results with observational data
for the reddening curves and the cosmic extinction.
Since there are large uncertainties in observational data,
we do not attempt any sophisticated fitting procedure, but
focus on presenting simple comparisons between our prediction and
the observations. However, even with the currently available
observational data, we are able to draw some interesting
conclusions as we see in this section.

\subsection{Reddening curve}\label{subsec:reddening_curve}

The dust grain size is effectively constrained by the reddening curve.
Following the assumption in MF12, we trace the reddening
in galaxy halos by Mg \textsc{ii} absorbers (see also the Introduction).
We use the
reddening curve derived from the reddening analysis of
background QSOs for Mg \textsc{ii} absorbers at $z\sim 1$ and 2 by MF12.
We adopt the SDSS $u$, $g$, $r$, and $i$ band data, and
do not use shorter Galaxy Evolution Explorer
(GALEX; \citealt{Martin:2005aa}) bands, where hydrogen, not dust,
dominates the absorption (see fig.\ 4 in MF12).

As mentioned in Section~\ref{subsec:model_reddening},
we adopt $N_\mathrm{H}=10^{19.5}$~cm$^{-2}$ for the fiducial value,
but considering the uncertainty in $N_\mathrm{H}$ and $\mathcal{D}$,
we examine an order of magnitude variation of the reddening by
changing $N_\mathrm{H}$ between $10^{19}$ and $10^{20}$~cm$^{-2}$
(but fixing $\mathcal{D}=0.006$ because of the degeneracy between
$N_\mathrm{H}$ and $\mathcal{D}$). The uncertainty given here
is more conservative {than} that given in MF12.

\begin{figure}
\includegraphics[width=0.45\textwidth]{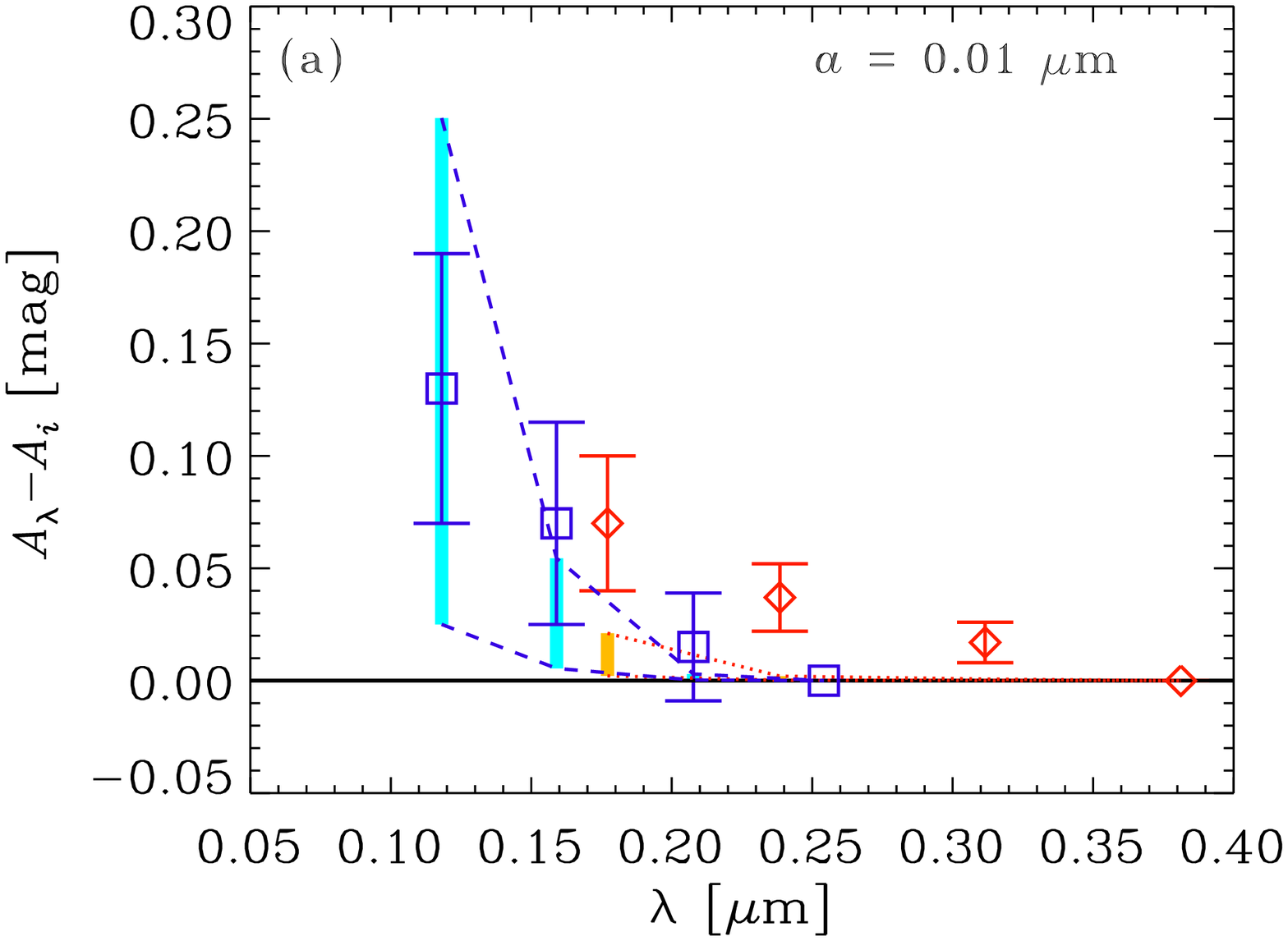}
\includegraphics[width=0.45\textwidth]{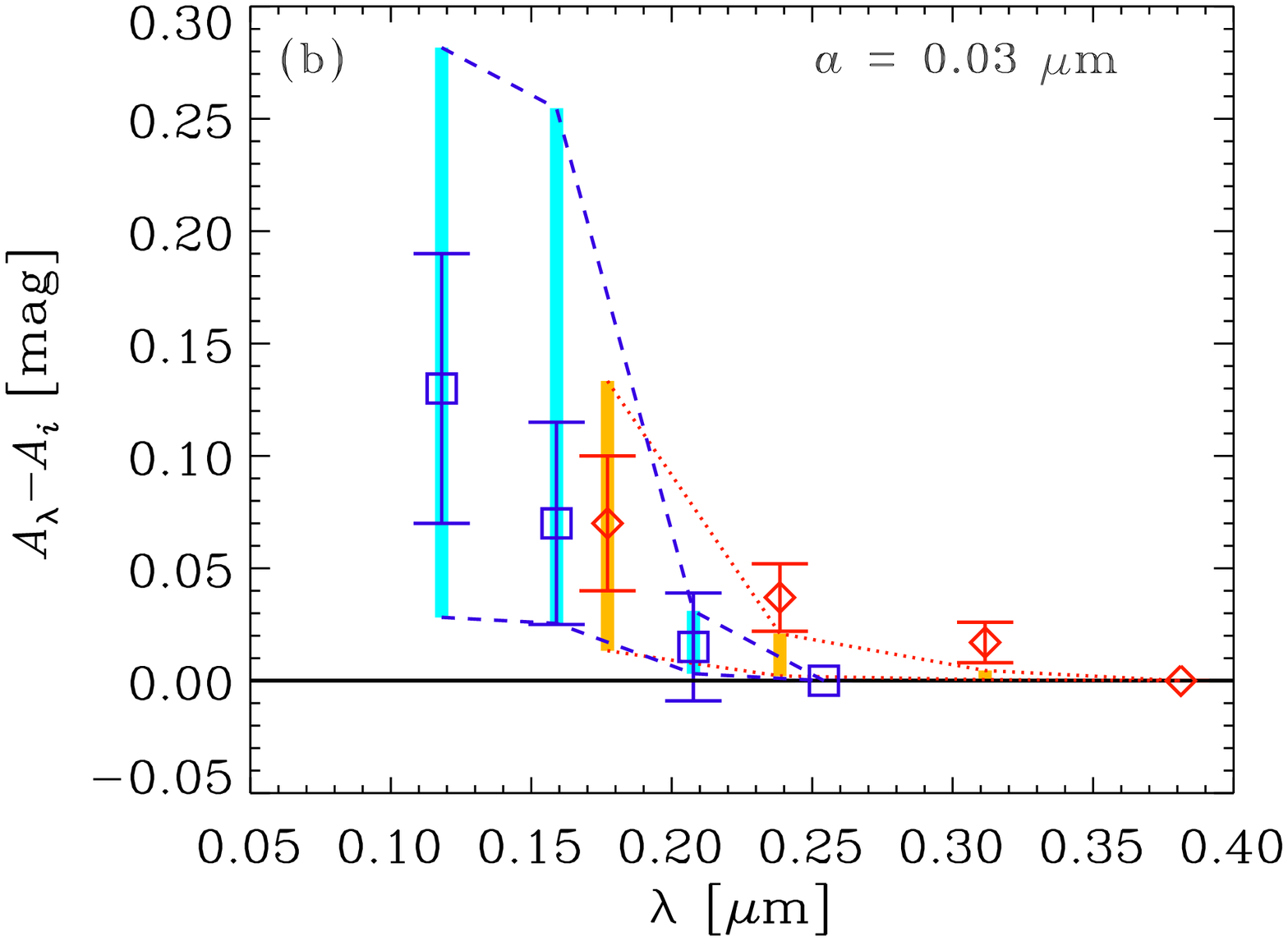}
\includegraphics[width=0.45\textwidth]{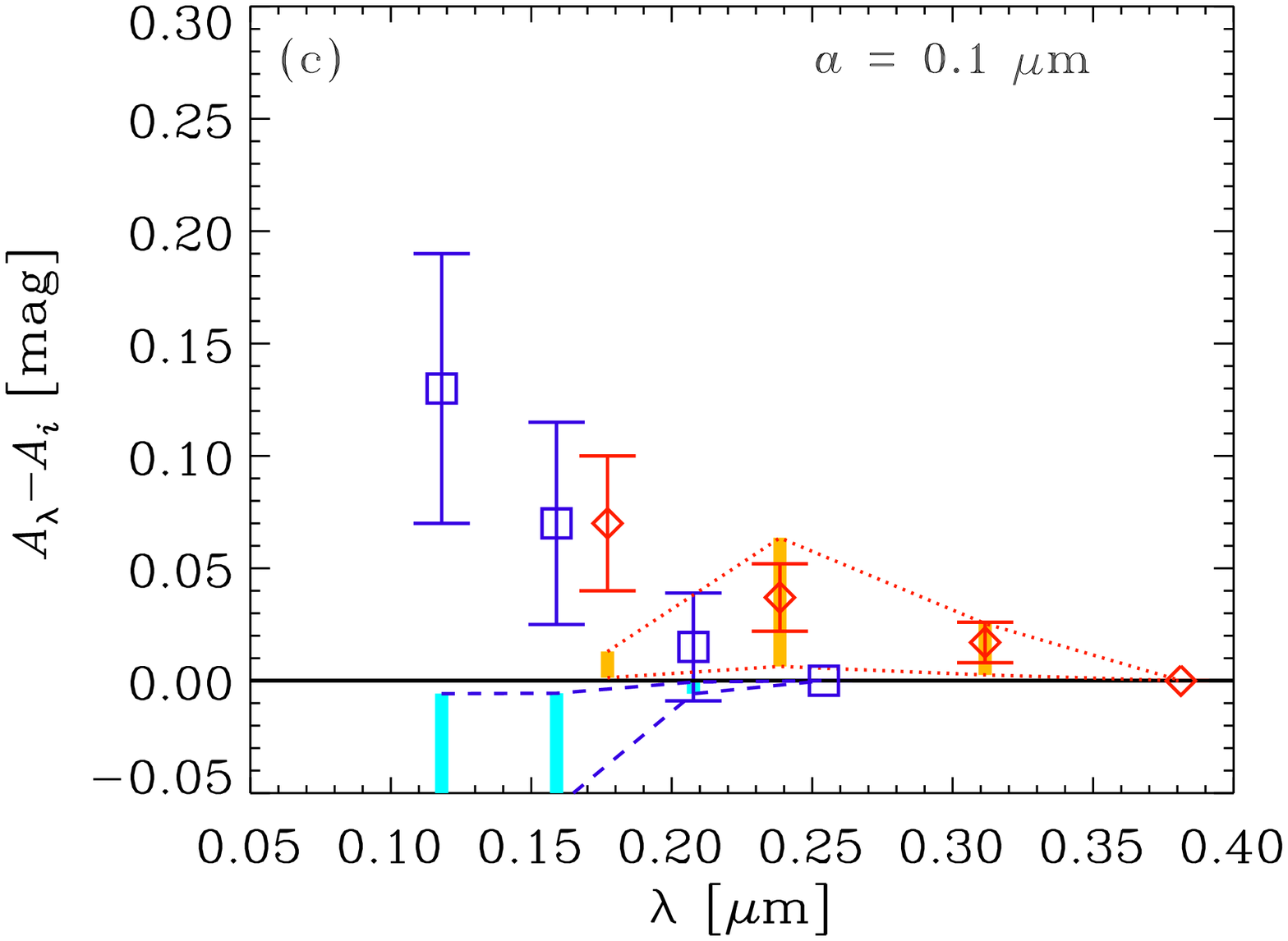}
\caption{Reddening curves for silicate with grain radii
(a) $a=0.01~\mu$m,
(b) $a=0.03~\mu$m, and
(c) $a=0.1~\mu$m at $z=1$ and 2.
{The orange and light blue bars show the theoretically predicted ranges
of $A_\lambda -A_i$ in the SDSS $u$, $g$, and $r$ bands
(the rest-frame wavelengths
are shown) at $z=1$ and 2, respectively. The red dotted and blue dashed lines connect the
upper and lower ranges corresponding to $N_\mathrm{H}=10^{19}$--$10^{20}$
cm$^{-2}$ at $z=1$ and 2, respectively.
The red diamonds and blue squares with error bars
show the observed reddening data for Mg \textsc{ii} absorbers at
$z=1$ and 2, respectively, taken from MF12. The error bars are expanded
by a factor of 3 relative to those in MF12 for a conservative comparison.
If the observational data overlap with the orange and light blue bars,
the model is successful.}
For $a=0.01~\mu$m, the reddening at $\lambda >0.25~\mu$m is too small to
be clearly seen.
\label{fig:reddening_sil}}
\end{figure}

In Fig.\ \ref{fig:reddening_sil}, we show the reddening, $A_\lambda -A_i$, as a function of
rest-frame wavelength $\lambda$ for silicate. We examine three cases for the grain
radius: $a=0.01$, 0.03, and 0.1~$\mu$m. We find that the observed
reddening curves are broadly reproduced with $a=0.03~\mu$m.
If the grain radius is as large as $a=0.1~\mu$m, the calculated
reddening curve is too flat (or $|A_\lambda -A_i|$ is too small) to
reproduce the data, which we confirmed to be true also for
$a>0.1~\mu$m.
If the grain radius is as small as $a=0.01~\mu$m, the calculated reddening
is too small (especially at $\lambda > 0.25~\mu$m), simply because
$A_\lambda$ is small in the relevant wavelength range (this is true also
for $a<0.01~\mu$m).
Therefore, the grain radii in Mg\,\textsc{ii} absorbers (and probably in
galaxy halos) are around 0.03~$\mu$m if the main composition is
silicate. Silicate grains larger than 0.1~$\mu$m or
smaller than 0.01~$\mu$m are not favored in explaining the reddening
curve of Mg \textsc{ii} absorbers.

\begin{figure}
\includegraphics[width=0.45\textwidth]{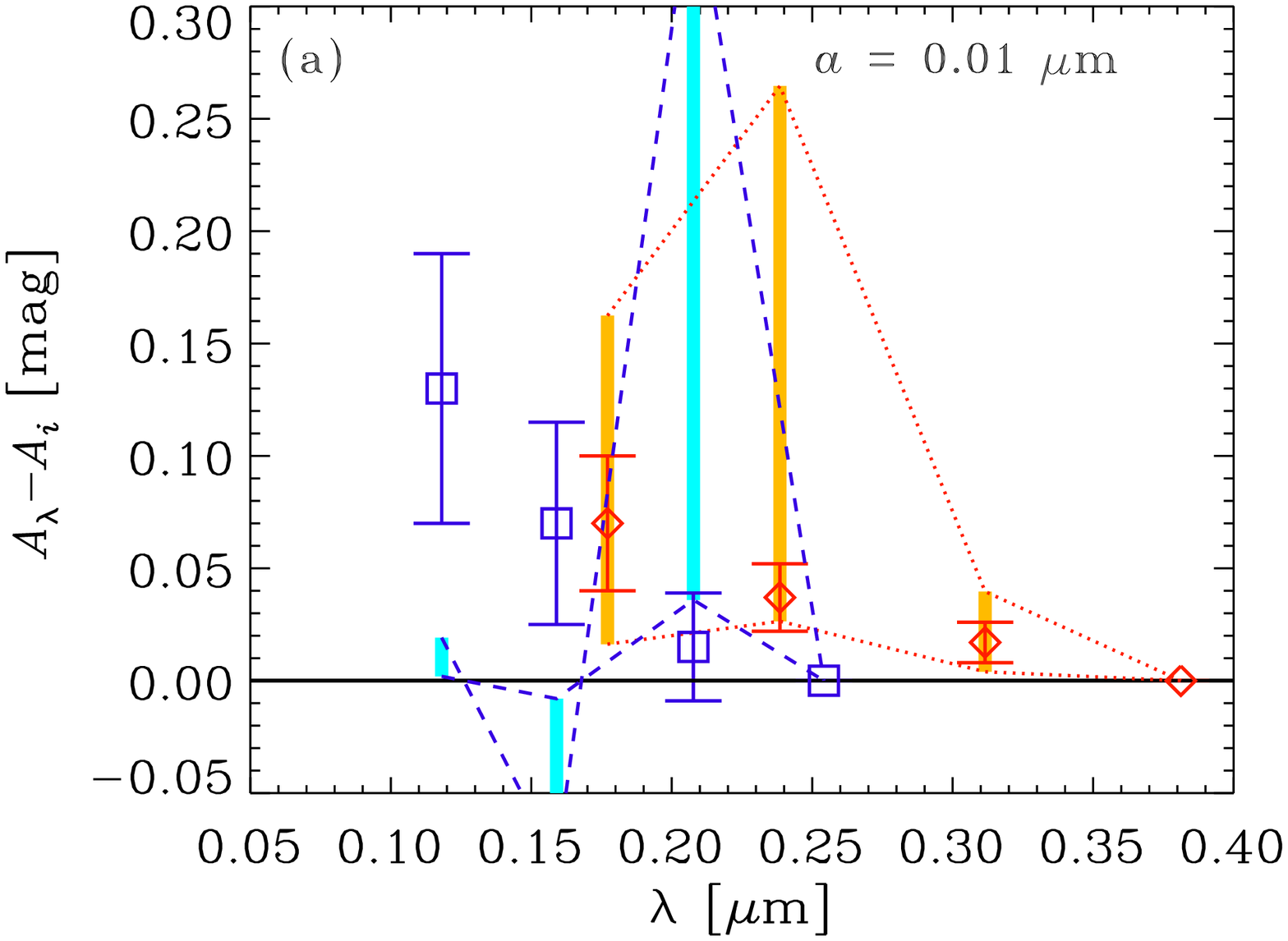}
\includegraphics[width=0.45\textwidth]{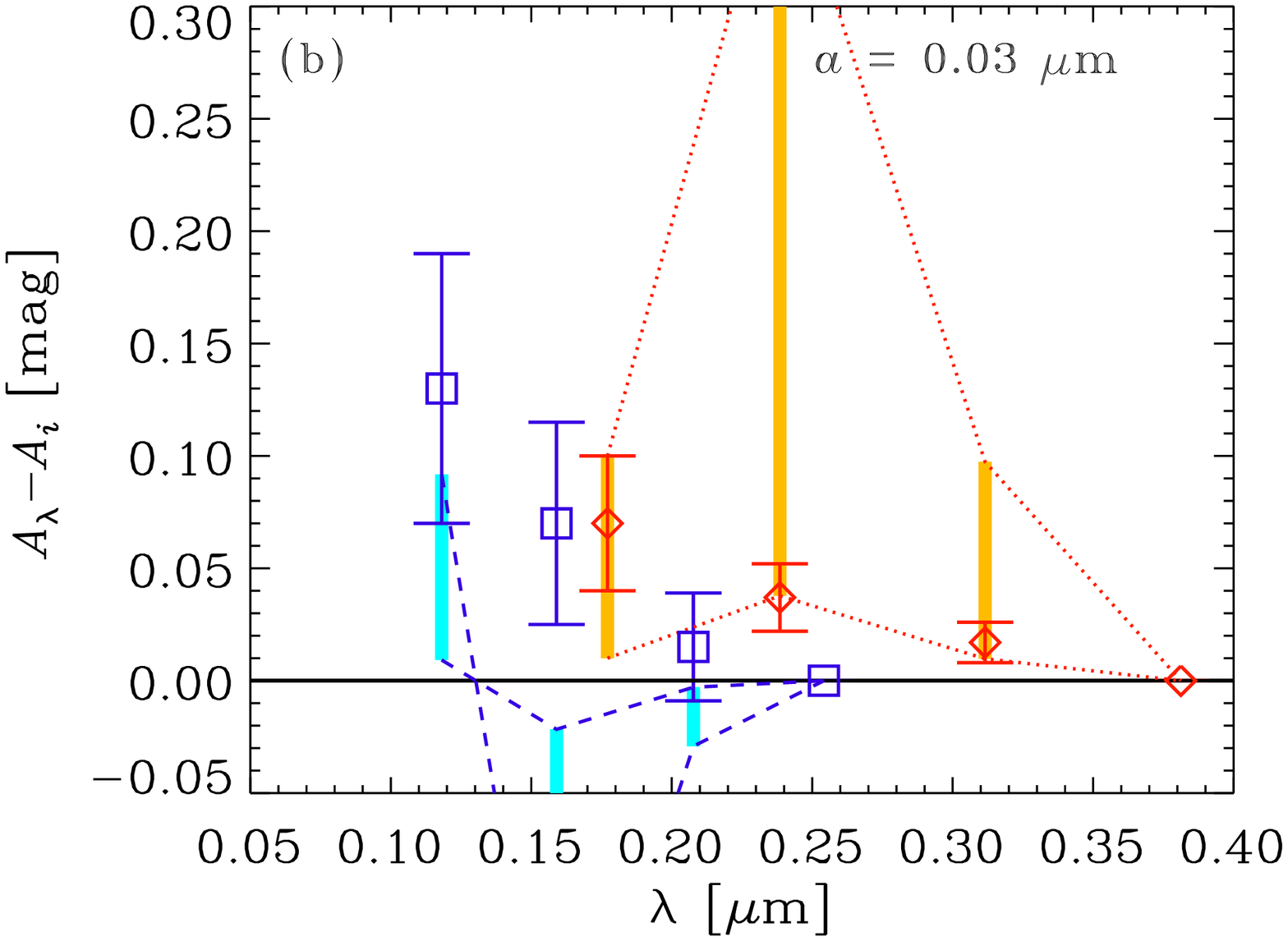}
\includegraphics[width=0.45\textwidth]{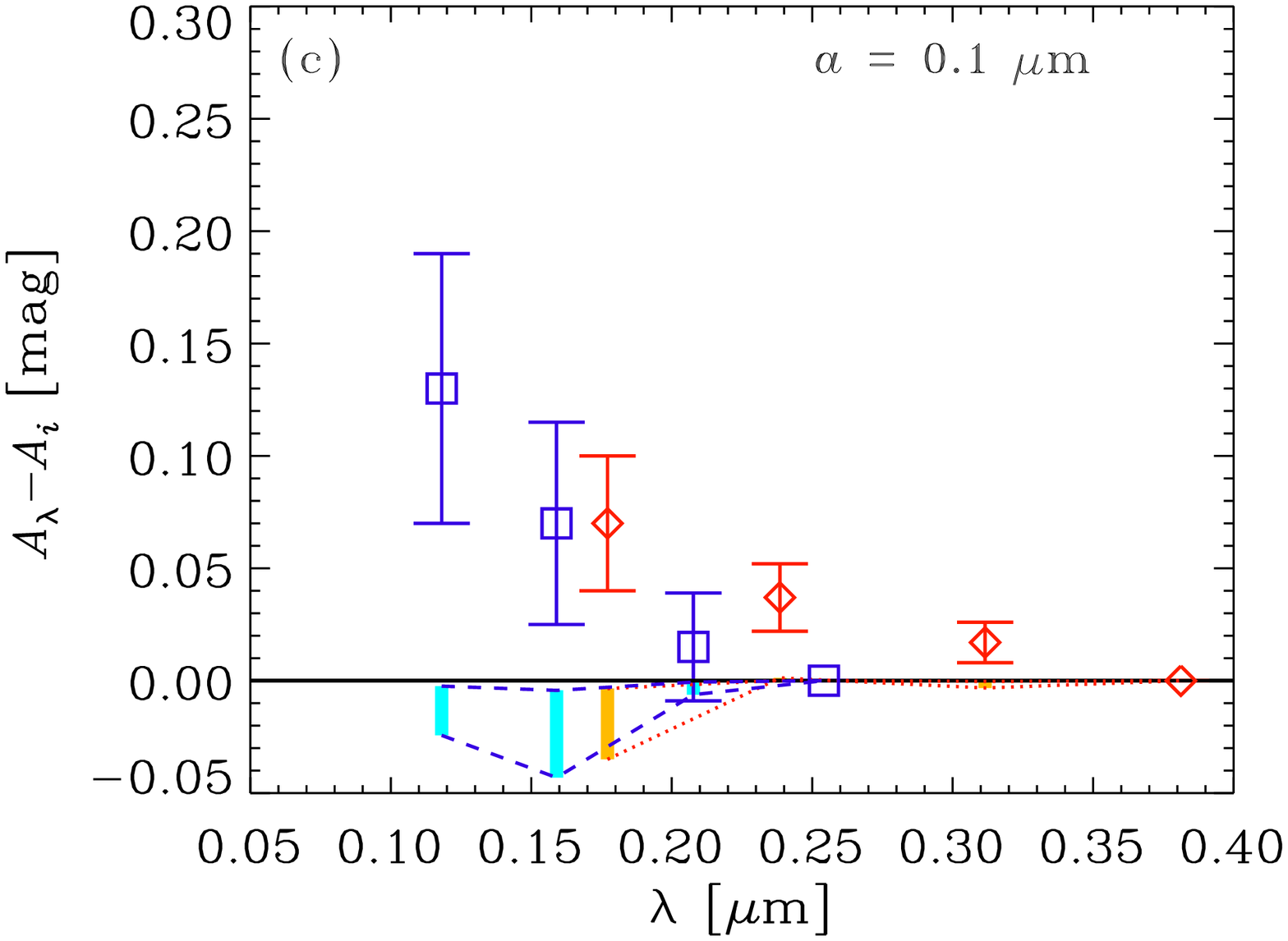}
\caption{Same as Fig.\ \ref{fig:reddening_sil} but for graphite.
\label{fig:reddening_gra}}
\end{figure}

In Fig.\ \ref{fig:reddening_gra}, we show the reddening curves for graphite.
We observe that, because of the prominent 2175 \AA\ bump of
small ($a\lesssim 0.03~\mu$m) graphite grains, the observed
reddening is difficult to reproduce with graphite.
In particular, the strongly non-monotonic behavior along the wavelength
is not supported by the observational data.
{This is why previous studies such as M10 and MF12 used the SMC extinction
curve, which is bumpless, to interpret the extinctions in galaxy halos
and Mg \textsc{ii} absorbers.}
For $a=0.1~\mu$m, the calculated reddening curve is
too flat as was also the case with silicate (this is also true for $a>0.1~\mu$m).
Therefore, graphite is not supported by the observed reddening curves.

\begin{figure}
\includegraphics[width=0.45\textwidth]{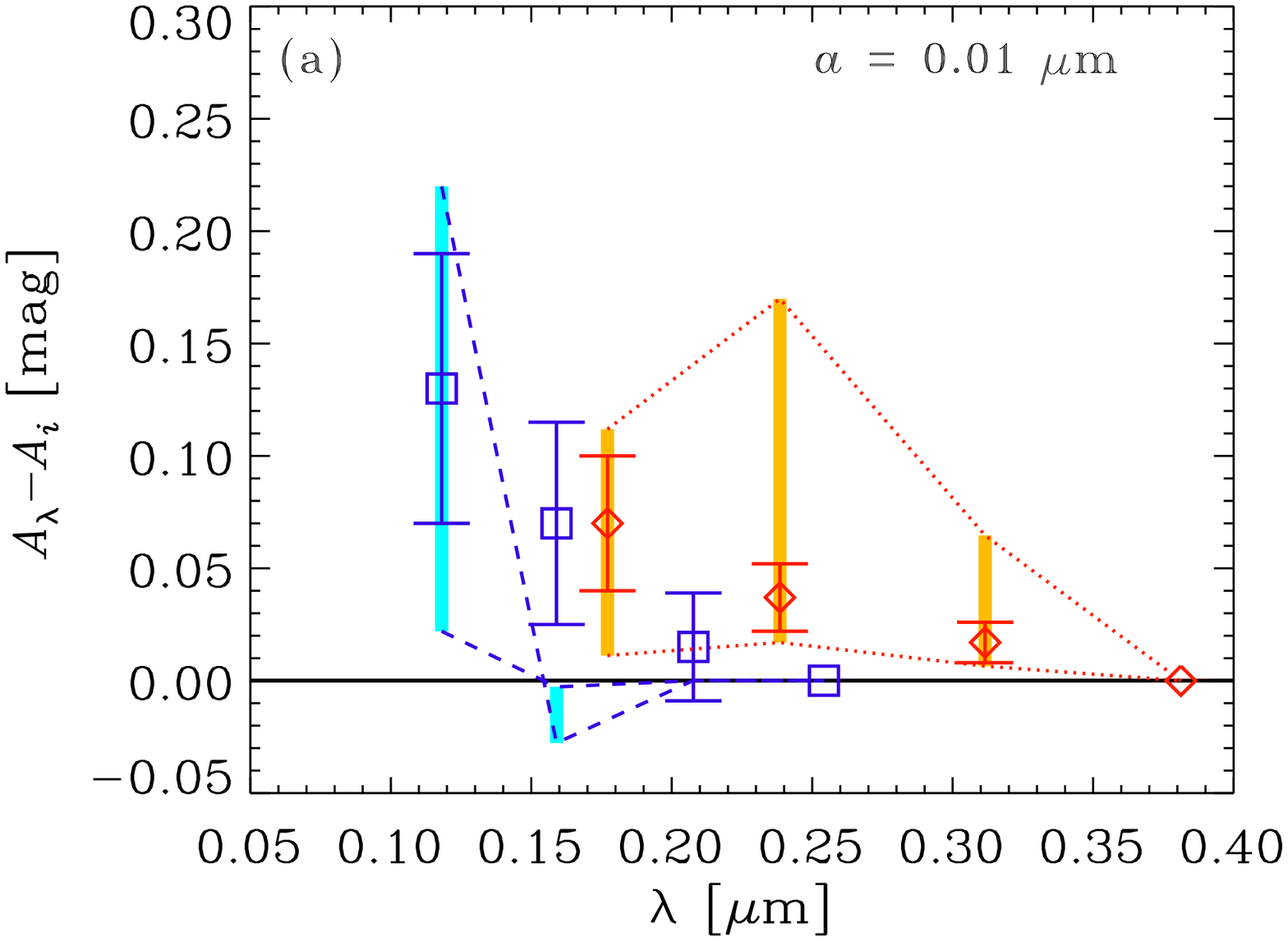}
\includegraphics[width=0.45\textwidth]{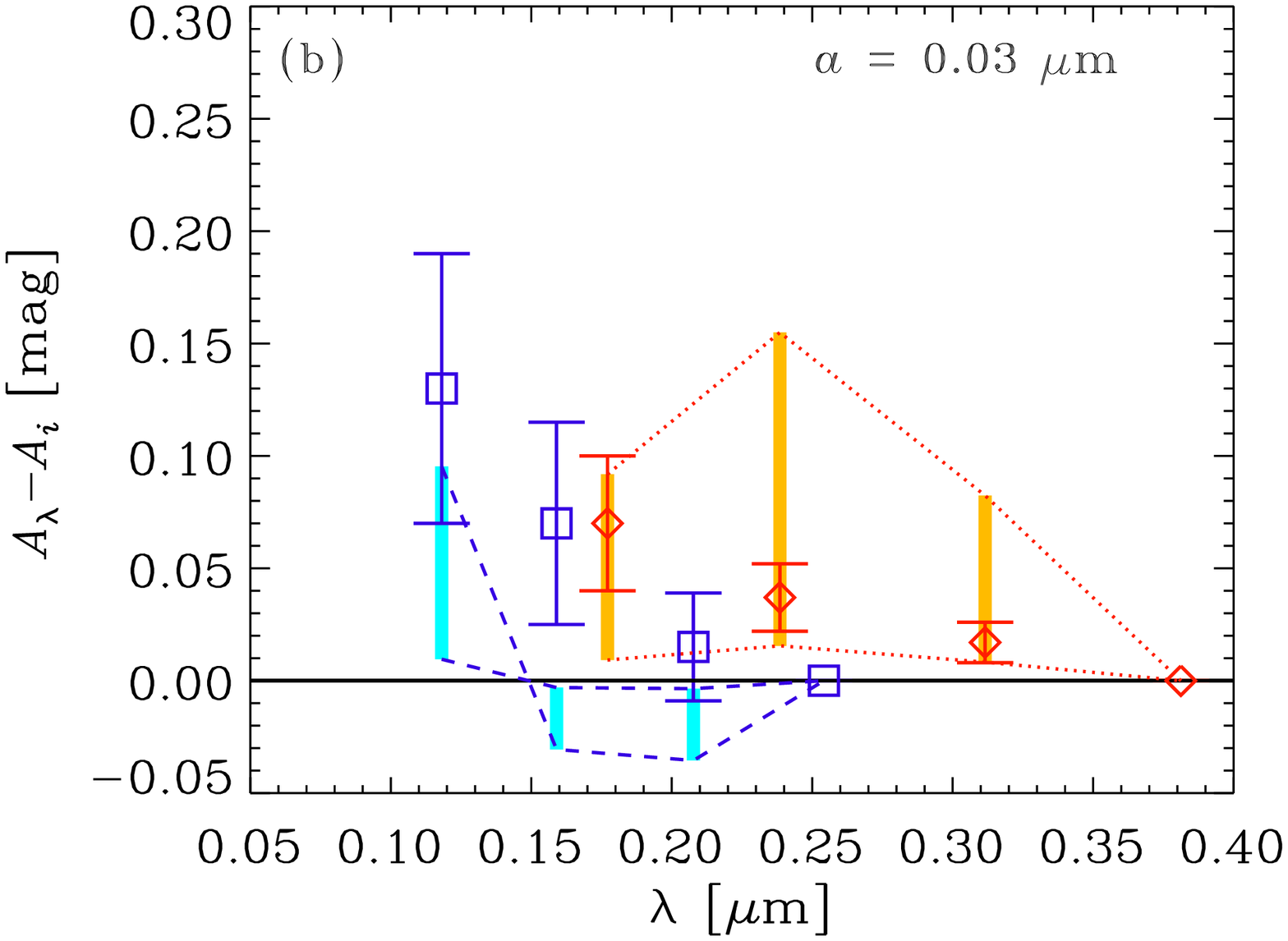}
\includegraphics[width=0.45\textwidth]{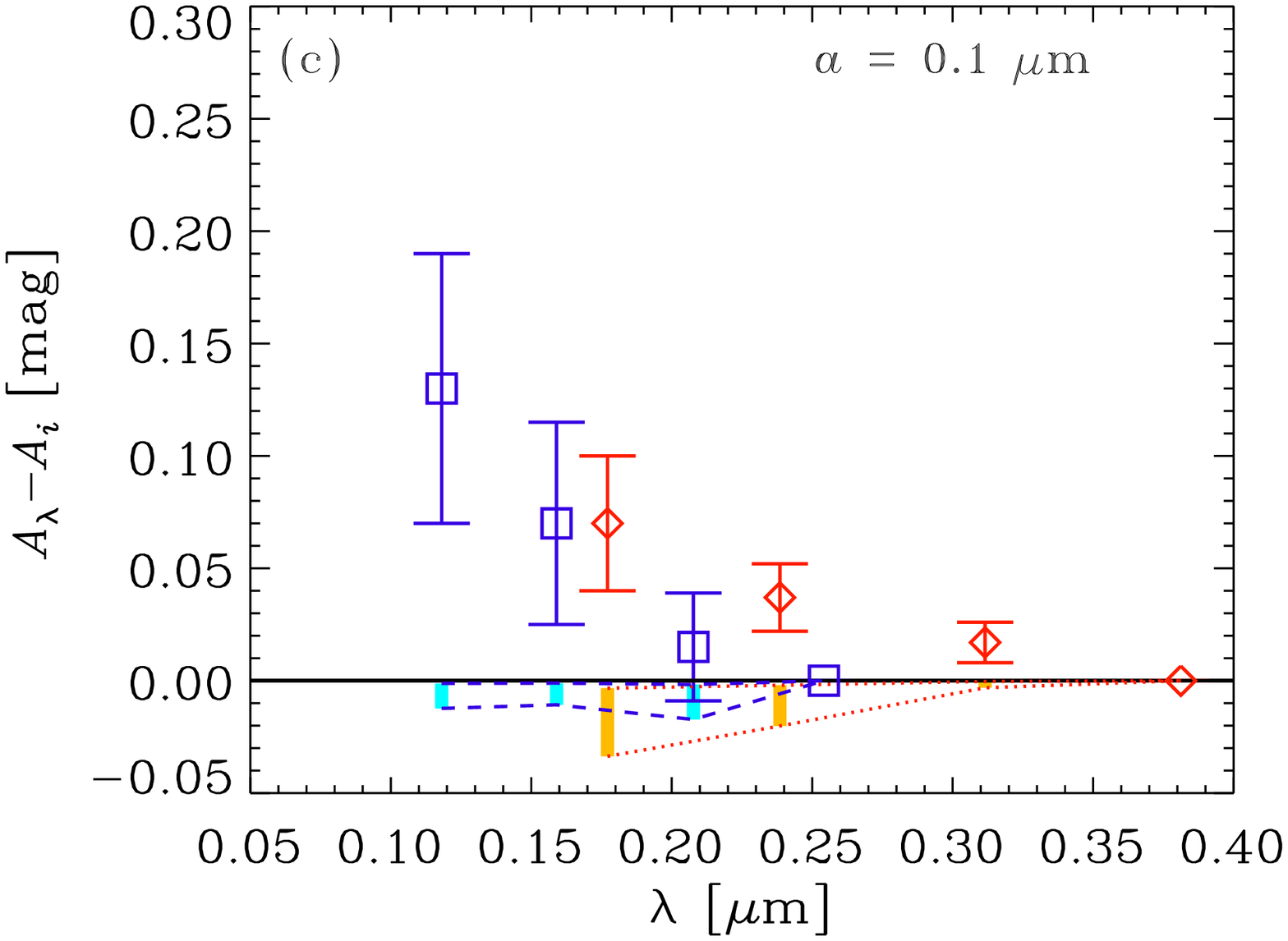}
\caption{Same as Fig.\ \ref{fig:reddening_sil} but for {amorphous carbon}.
\label{fig:reddening_AC}}
\end{figure}

Another carbonaceous species, amorphous carbon, does not have such a
strong feature; thus, we expect that the observed reddening curves are
better explained by amorphous carbon than by graphite.
In Fig.\ \ref{fig:reddening_AC}, we show the reddening curves calculated
for amorphous carbon.
Because of the weaker bump, as expected, amorphous carbon fits
the observed reddening curves better than graphite
for $a=0.01$ and 0.03~$\mu$m.
For $a=0.1~\mu$m, the calculated reddening curve is
too flat, as was also the case with graphite.
Therefore, a carbonaceous material with a weak feature
is marginally accepted by the observed reddening curves if the
grain radius is $a\lesssim 0.03~\mu$m.

\subsection{Cosmic extinction}\label{subsec:result_ext}

The second constraint on the dust properties is obtained
from the cosmic extinction.
For comparison with observational data, we adopt the upper and
lower limits for the cosmic extinction at $z<2$ compiled in M10.
We refer to the appendix in M10 and the references therein
for the summary of those data, and
only briefly describe them in what follows.
\citet{Avgoustidis:2009aa} obtained a constraint on the dust opacity
up to $z\sim 0.5$ by combining the data of the baryon acoustic oscillation and
Type Ia SNe. They obtained an upper limit of
extinction at the visible wavelength as 0.08 mag.
\citet{Mortsell:2003aa} analyzed QSO colors at $0.5<z<2$,
and put an upper limit of $A_V<0.2$~mag at $z\simeq 1$.
\citet{Menard:2008aa} measured the reddening of Mg \textsc{ii} absorbers,
which are interpreted as residing in halos of $\sim L^*$ galaxies.
Based on their statistical data, M10 estimated the contribution
of Mg \textsc{ii} absorbers to the cosmic extinction, obtaining
$A_V>0.009$, $>0.029$, and $>0.044$~mag at $z=0.6$, 1.3 and 1.8,
respectively. Those $A_V$ values are lower limits because they only count
Mg \textsc{ii} absorbers. Strictly speaking, their extinction estimates
assume that the SMC extinction curve is applicable, which may not
be consistent with our adopted wavelength dependence of $\kappa_\nu$.
We comment on this issue later in Section \ref{subsec:consistency}.

We calculate the cosmological extinction
$\tau_\mathrm{d,cosmo}(z,\,\lambda_\mathrm{obs})$ (see Section \ref{subsec:model_ext}).
Precisely speaking, the observer's frame wavelength $\lambda_\mathrm{obs}$ varies
slightly among
the observational constraints, but it is always around the $V$ band.
We simply represent $\lambda_\mathrm{obs}$ by
the $V$-band wavelength (0.55 $\mu$m) following M10.
This particular choice of wavelength does not affect our discussions below.
We present the extinction in units of magnitude by multiplying $\tau_\mathrm{d,cosmo}$
by $2.5\log \mathrm{e}=1.086$.

\begin{figure}
\begin{center}
\includegraphics[width=0.4\textwidth]{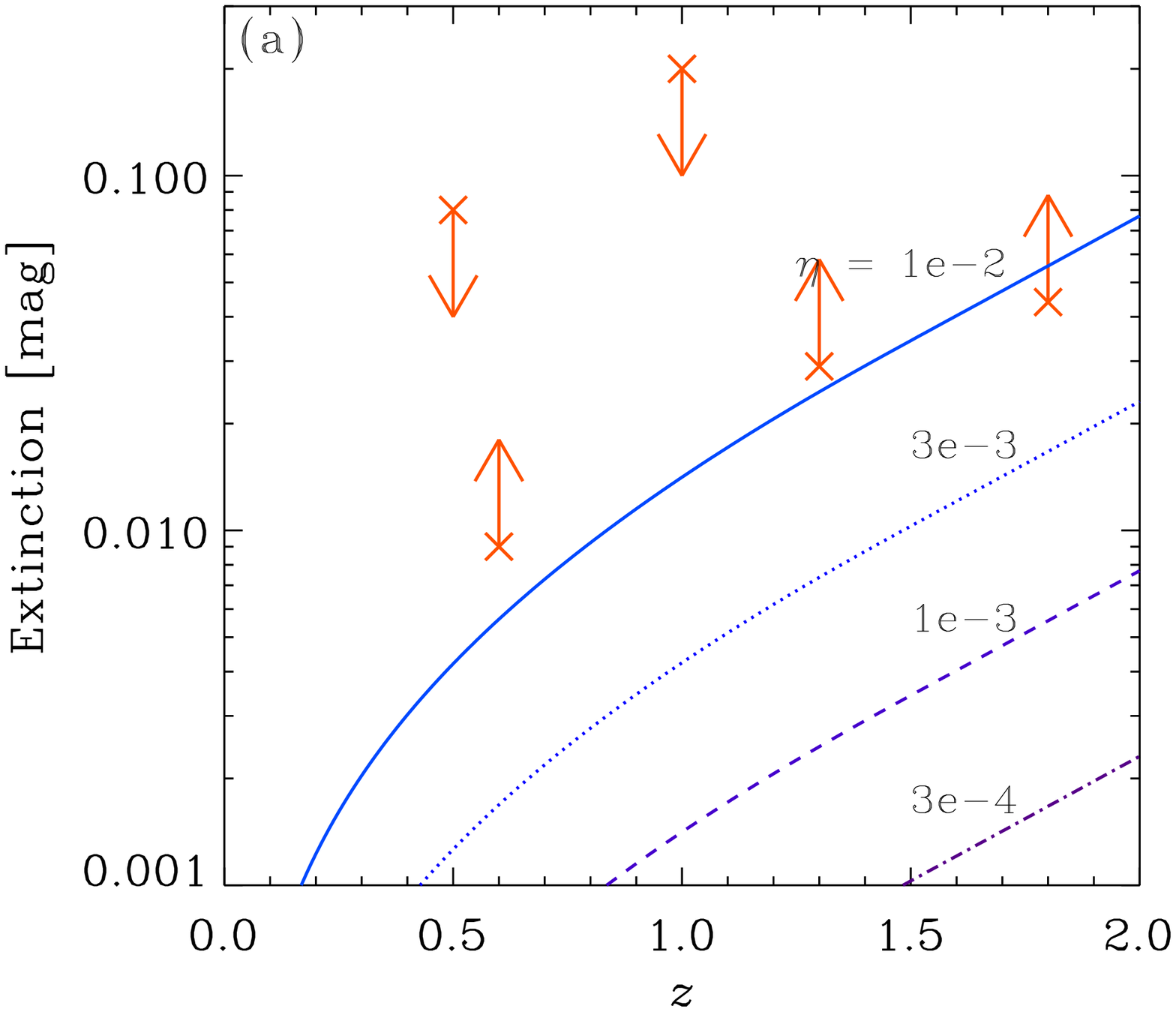}
\includegraphics[width=0.4\textwidth]{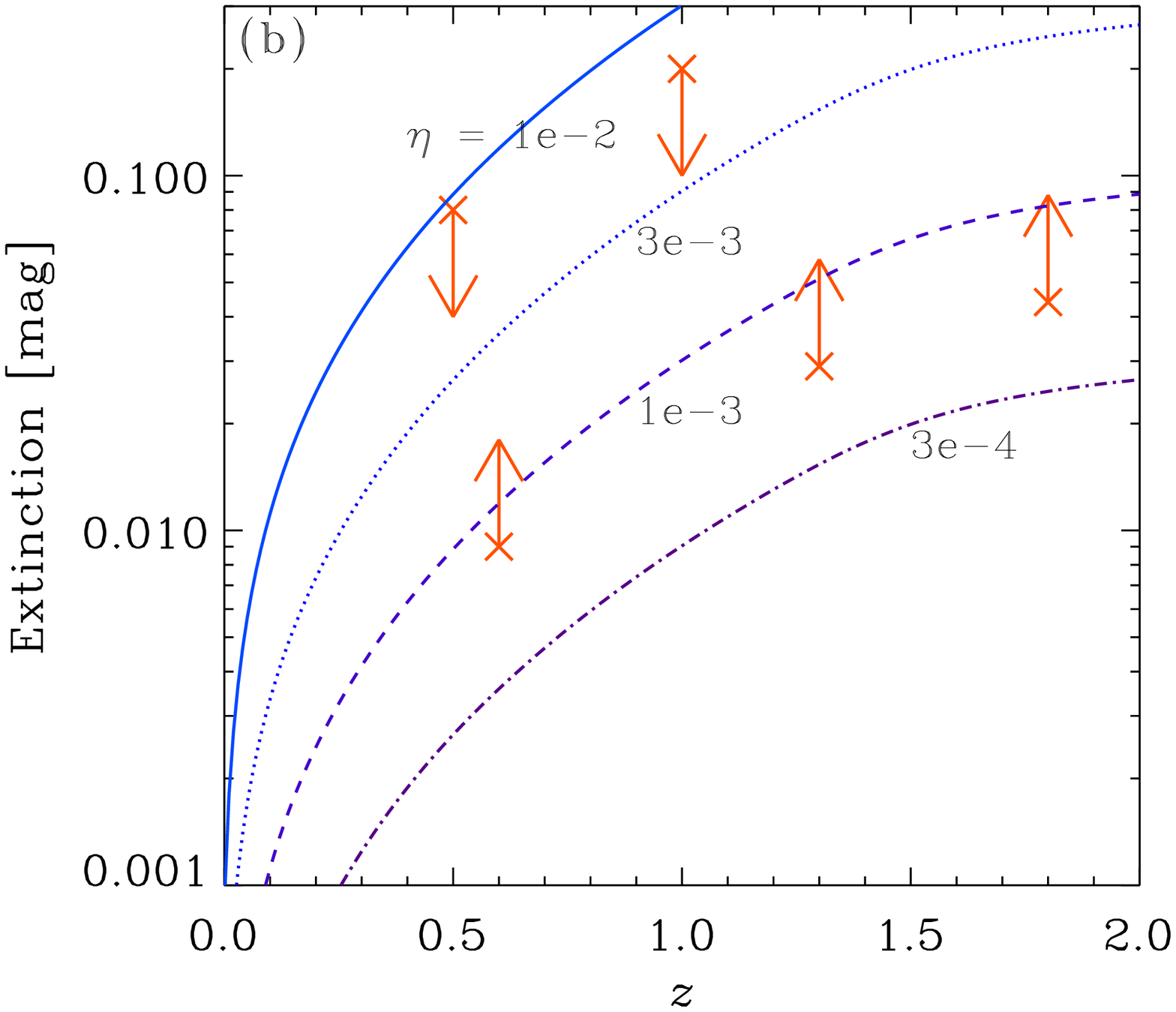}
\includegraphics[width=0.4\textwidth]{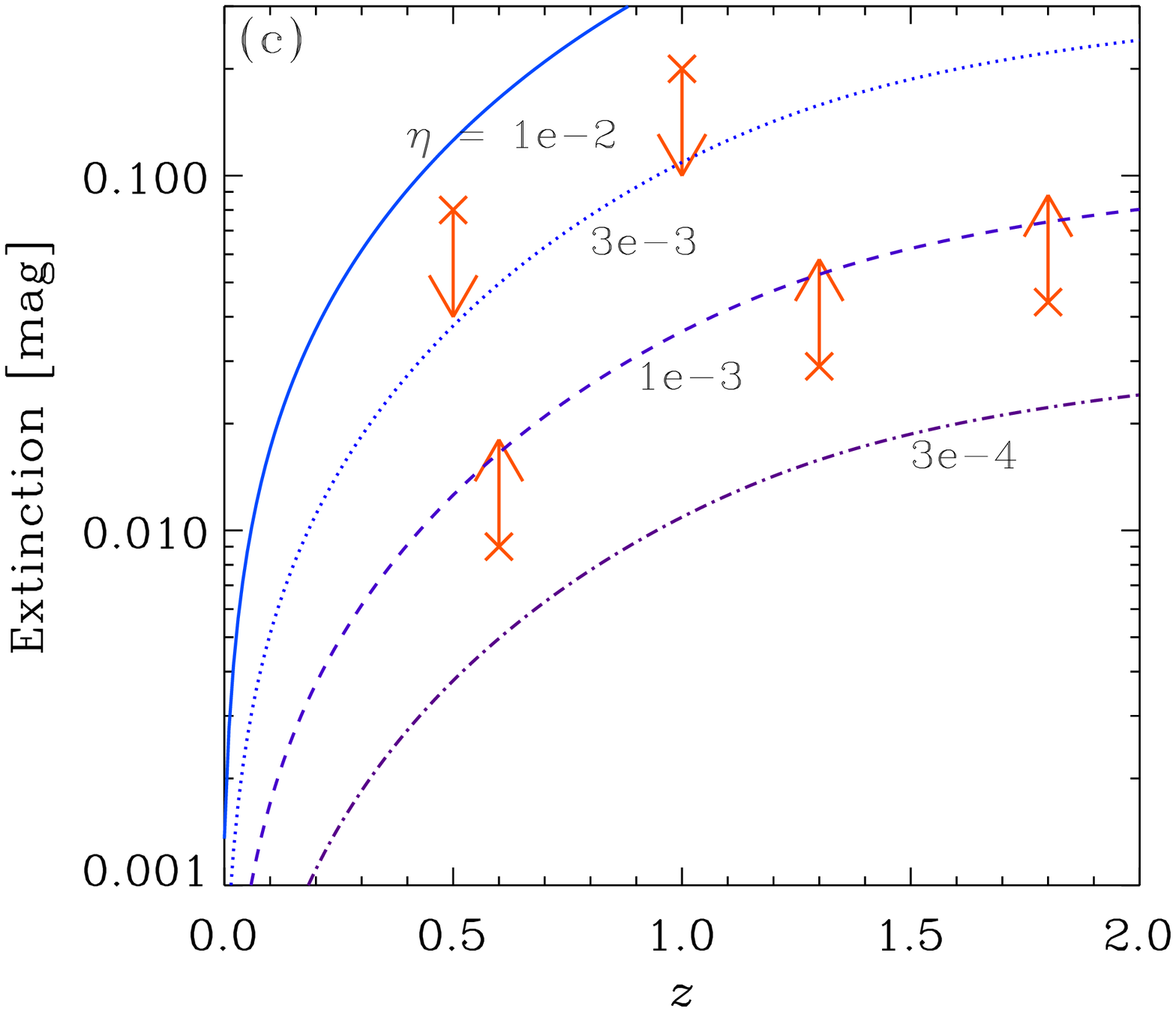}
\end{center}
\caption{Extinction as a function of redshift (cosmic extinction) for
various $\eta$
(dust abundance in the halo relative to the stellar mass) with
a fixed grain size of $a=0.03~\mu$m.
Panels (a), (b), and (c) show silicate, graphite and amorphous carbon.
{The solid, dotted, dashed, and dot-dashed lines present the results for
$\eta=10^{-2}$, $3\times 10^{-3}$, $10^{-3}$, and $3\times 10^{-4}$,
respectively.} Upper (downward arrows) and lower (upward arrows)
limits at various redshifts are taken from M10 (see the references therein).
\label{fig:tau_eta}}
\end{figure}

First, we show the results for various $\eta$ with a fixed radius $a=0.03~\mu$m
(this grain size is favored by the reddening curve as shown above)
in Fig.\ \ref{fig:tau_eta}. In this case, the cosmic extinction is simply
proportional to $\eta$. We observe that, for consistency with the observational
constraints, $\eta > 10^{-2}$ is required for silicate,
while the fiducial value $\eta\sim 10^{-3}$
as suggested above is consistent with the observational constraints
for graphite and amorphous carbon.

\begin{figure}
\begin{center}
\includegraphics[width=0.4\textwidth]{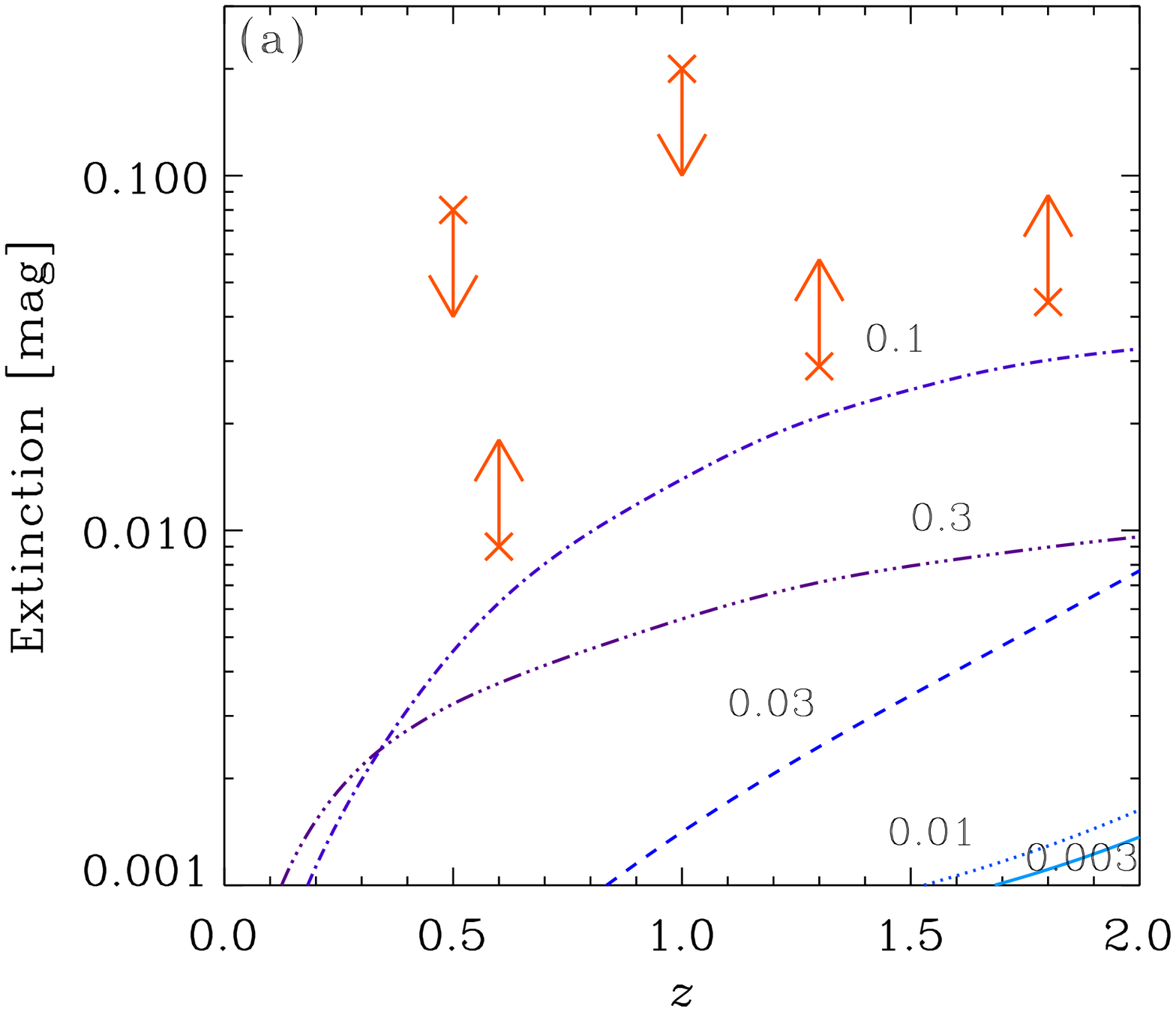}
\includegraphics[width=0.4\textwidth]{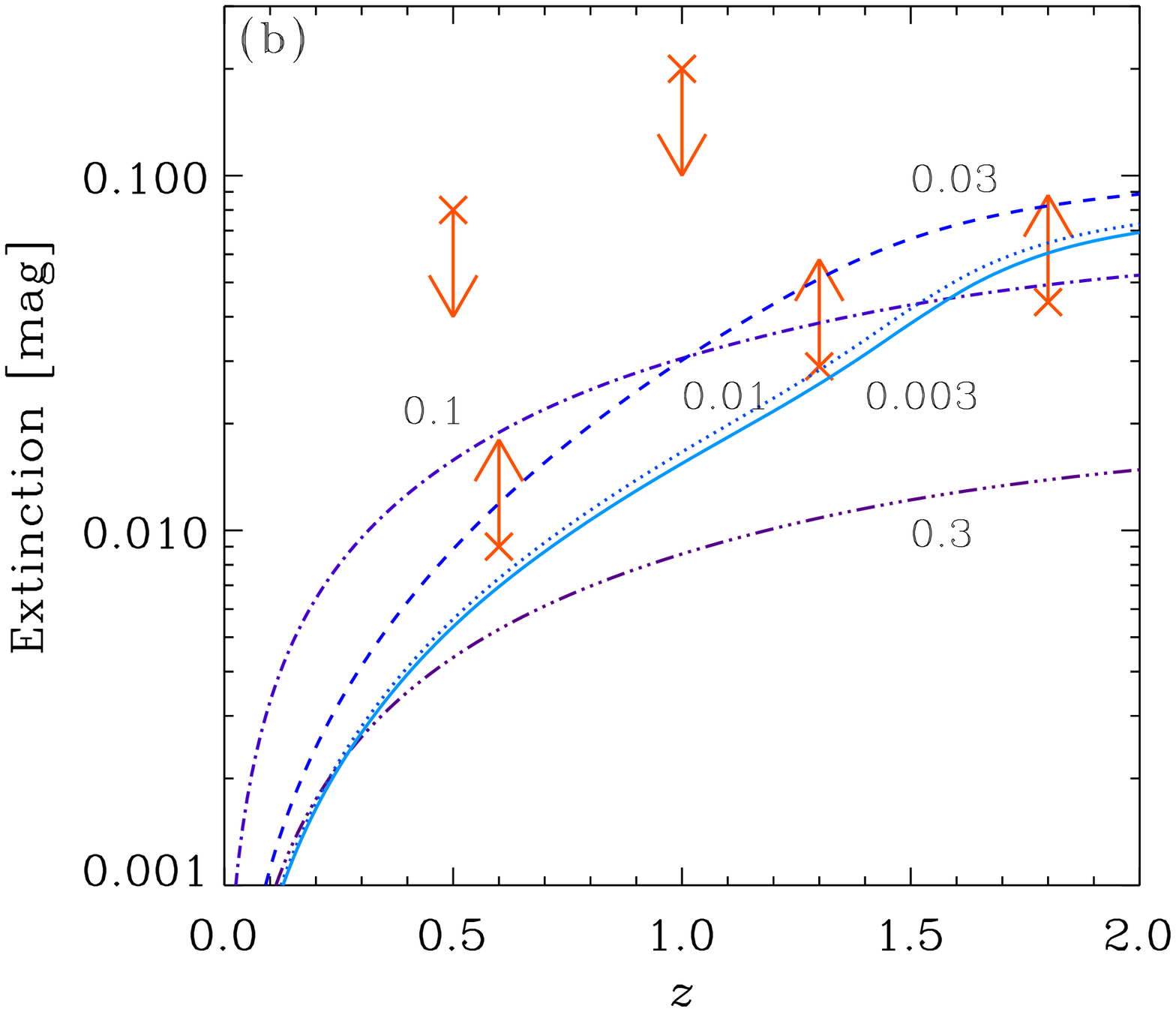}
\includegraphics[width=0.4\textwidth]{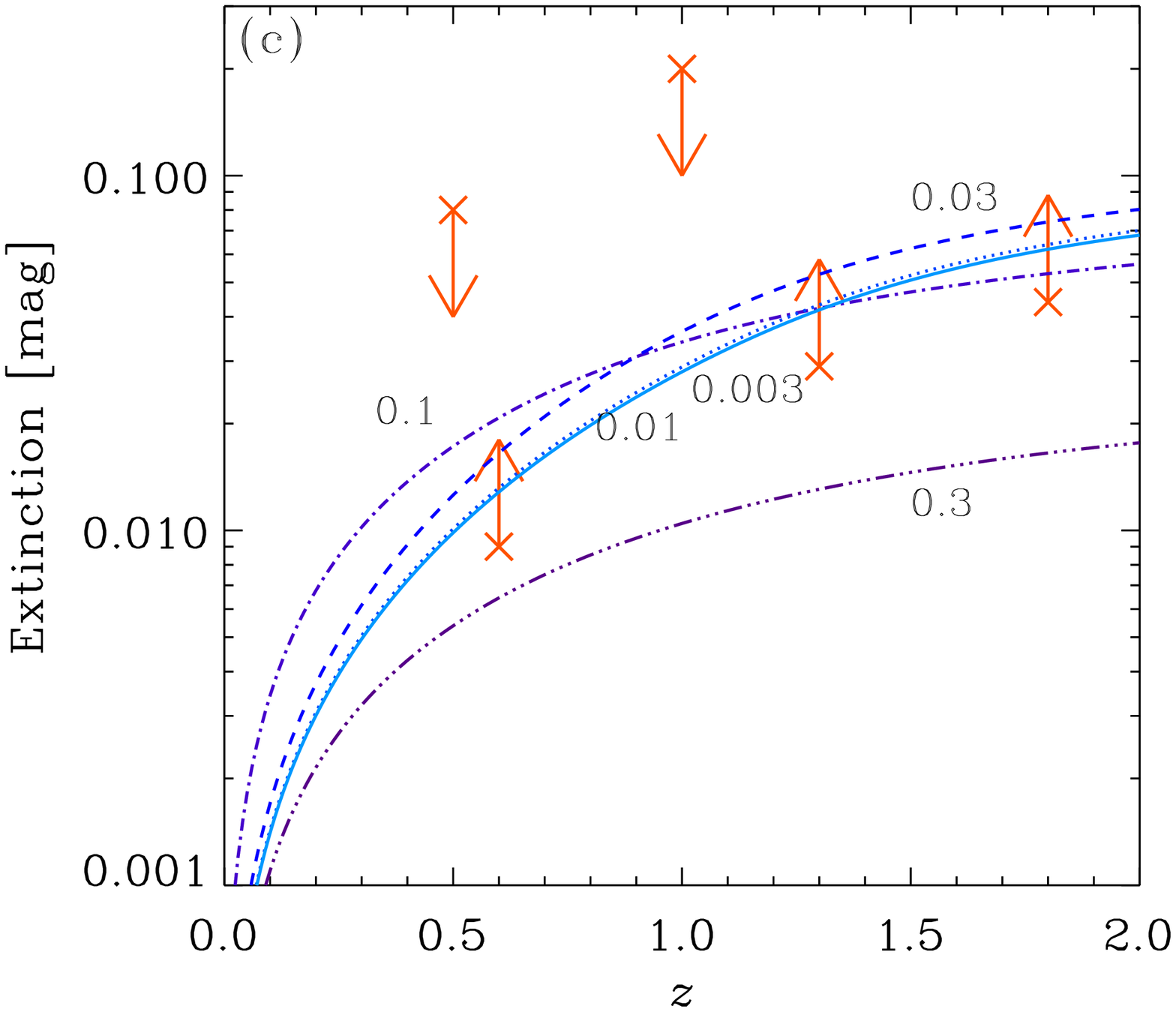}
\end{center}
\caption{Same as Fig.\ \ref{fig:tau_eta} but for various grain radii
{with a fixed $\eta =10^{-3}$. The solid, dotted, dashed,
dot-dashed, and dot-dot-dot-dashed lines show the results for
$a=0.003$, 0.01, 0.03, 0.1, and 0.3 $\mu$m, respectively.}
\label{fig:tau_size}}
\end{figure}

Next, we calculate the cosmological extinction
for various grain radii with a fixed
$\eta =10^{-3}$ in Fig.\ \ref{fig:tau_size}. As we already
discussed above, the extinction of silicate is not large enough to
satisfy the observational constraints with $\eta =10^{-3}$.
{We also find that
the silicate extinction is sensitive to the grain radius:
for $a=0.1~\mu$m, $\eta$ slightly larger than $10^{-3}$ is consistent
with the observational constraints. Thus, large grains
with $a\sim 0.1~\mu$m is favored by the cosmic extinction if the
main dust composition is silicate and $\eta\sim 10^{-3}$.}
The extinctions of graphite and amorphous carbon are less sensitive to the
grain radius. With the fiducial value of $\eta$ ($=10^{-3}$),
graphite and amorphous carbon predict cosmic extinctions consistent with
the observational constraints {unless the grain
radius is $\gtrsim 0.3~\mu$m.}

\section{Discussion}\label{sec:discussion}

\subsection{Consistency between reddening and cosmic extinction}
\label{subsec:consistency}

We have used two observational constraints: one is the reddening curves
observed for Mg \textsc{ii} absorbers (assuming that
Mg\,\textsc{ii} absorbers trace the medium in galaxy halos),
and the other is the cosmic
extinction. Below, we summarize the constraint on the grain radius obtained
for each grain
species in Section \ref{sec:result}, and examine if we find grain radii that explain both of the above
observational constraints.

For silicate, the reddening curves favor $a\sim 0.03~\mu$m, while
the cosmic extinction indicates that we need a halo dust abundance
more than {$\sim 10$ times higher than suggested by observation
(i.e., $\eta\gtrsim 10^{-2}$)} if $a\sim 0.03~\mu$m.
The cosmic extinction is marginally consistent with $\eta\sim 10^{-3}$ if
{$a\sim 0.1~\mu$m. Such a large grain size} predicts too small a
reddening as shown in Fig.\ \ref{fig:reddening_sil}a.
Therefore, if silicate
is the main composition,
a large amount of dust $\eta >10^{-2}$ in galaxy halos
is required (in other words, we do not find a consistent grain radius
under $\eta =10^{-3}$).

For graphite, the observed reddening curves are not reproduced well
because of the prominent 2175 \AA\ bump. In contrast, the cosmic
extinction is consistent with the observational constrains in a
broad range of grain radius with the fiducial value of $\eta$ ($=10^{-3}$).
Thus, we tried a carbonaceous material without a strong bump in
the ultraviolet (UV); that is, amorphous carbon. For amorphous carbon, the
reddening curves prefer $a\lesssim 0.03~\mu$m, while the
cosmic extinction {favors $a\lesssim 0.1~\mu$m under
$\eta =10^{-3}$. Thus, small grains with $a\lesssim 0.03~\mu$m is
acceptable for both the reddening curves and the cosmic extinction.}

In summary, depending on the main dust composition,
there are two lines of ``solutions'' for the consistency
between the reddening curves and the cosmic extinction:
if the main material is
silicate, grains should have radii $\sim 0.03~\mu$m but
{a higher dust abundance ($\eta >10^{-2}$)} in
galaxy halos than suggested by the observations is required.
If the main component of dust in galaxy halos
is carbonaceous dust, it should not have a strong bump in the UV
and should have {grain radii $a\lesssim 0.03~\mu$m} under
the fiducial value of $\eta =10^{-3}$.

Precisely speaking, the observational lower limits on the cosmic
extinction in Figs.\ \ref{fig:tau_eta} and \ref{fig:tau_size}
have been derived using the SMC extinction
curve. If large grains ($a\gtrsim 0.1~\mu$m),
whose extinction curve is
significantly flatter than the SMC curve, are the main component
of dust in galaxy halos, more dust is necessary to reproduce
the same amount of reddening. That is, the observational
lower limits shown in Figs.\ \ref{fig:tau_eta} and \ref{fig:tau_size}
are raised if the grain size is large. Therefore,
the discrepancy between our theoretical predictions for
$a\gtrsim 0.1~\mu$m and the lower limits become larger
in Fig.\ \ref{fig:tau_size}a. Such large grains
are, however, not favored by the reddening constraints obtained in
Section \ref{subsec:reddening_curve}.

\subsection{Effects of grain size distribution}\label{subsec:sizedist}

{In the above, we have simply assumed that all the grains
have a single radius. This radius is interpreted as the typical radius of
grains which dominate the extinction opacity.
This approach is useful because constraining the grain
size distribution only with sparse wavelength sampling of
a reddening curve is a highly degenerate problem.
It is expected that
the above conclusion on the necessity of small ($a\lesssim 0.03~\mu$m)
grains in reproducing the reddening curve is robust even if we consider
a grain size distribution.}

{Although the full grain size distribution is impossible to determine,
it is worth examining the effect of grain size distribution on the conclusion.
In particular, we focus on examining if small
($a\lesssim 0.03~\mu$m) grains are required robustly.
To concentrate on the existence of small grains, we fix the maximum
grain radius, $a_\mathrm{max}$, and the functional form
of the grain size distribution but move the minimum grain radius, $a_\mathrm{min}$:
\begin{eqnarray}
n(a)=Ca^{-3.5}~~~(a_\mathrm{min}\leq a\leq a_\mathrm{max}),\label{eq:MRN}
\end{eqnarray}
where $C$ is the normalizing constant (but as we show below, $C$ is not
important). This power-law form is appropriate for the Milky Way dust
\citep[][hereafter MRN]{Mathis:1977aa}. Although there is no physical reason that this functional
form is applicable to the IGM/CGM dust, we simply adopt this form to examine the
effect of small grains ($a_\mathrm{min}$).
We adopt $a_\mathrm{max}=0.25~\mu$m according to MRN, while
we vary $a_\mathrm{min}=0.01$, 0.03, and 0.1 $\mu$m.}

{We extend the formulation in Section \ref{subsec:model_reddening}
by considering the grain size distribution. The mass extinction coefficient
is expressed instead of equation (\ref{eq:kappa}) as
\begin{eqnarray}
\kappa_\mathrm{ext}(\lambda )=
\frac{\int_{a_\mathrm{min}}^{a_\mathrm{max}}\pi a^2Q_\mathrm{ext}(\lambda ,\, a)n(a)\,da}
{\int_{a_\mathrm{min}}^{a_\mathrm{max}}\frac{4}{3}\pi a^3sn(a)\, da}.
\end{eqnarray}
Here the above normalization constant $C$ cancels out.
}

\begin{figure}
\includegraphics[width=0.45\textwidth]{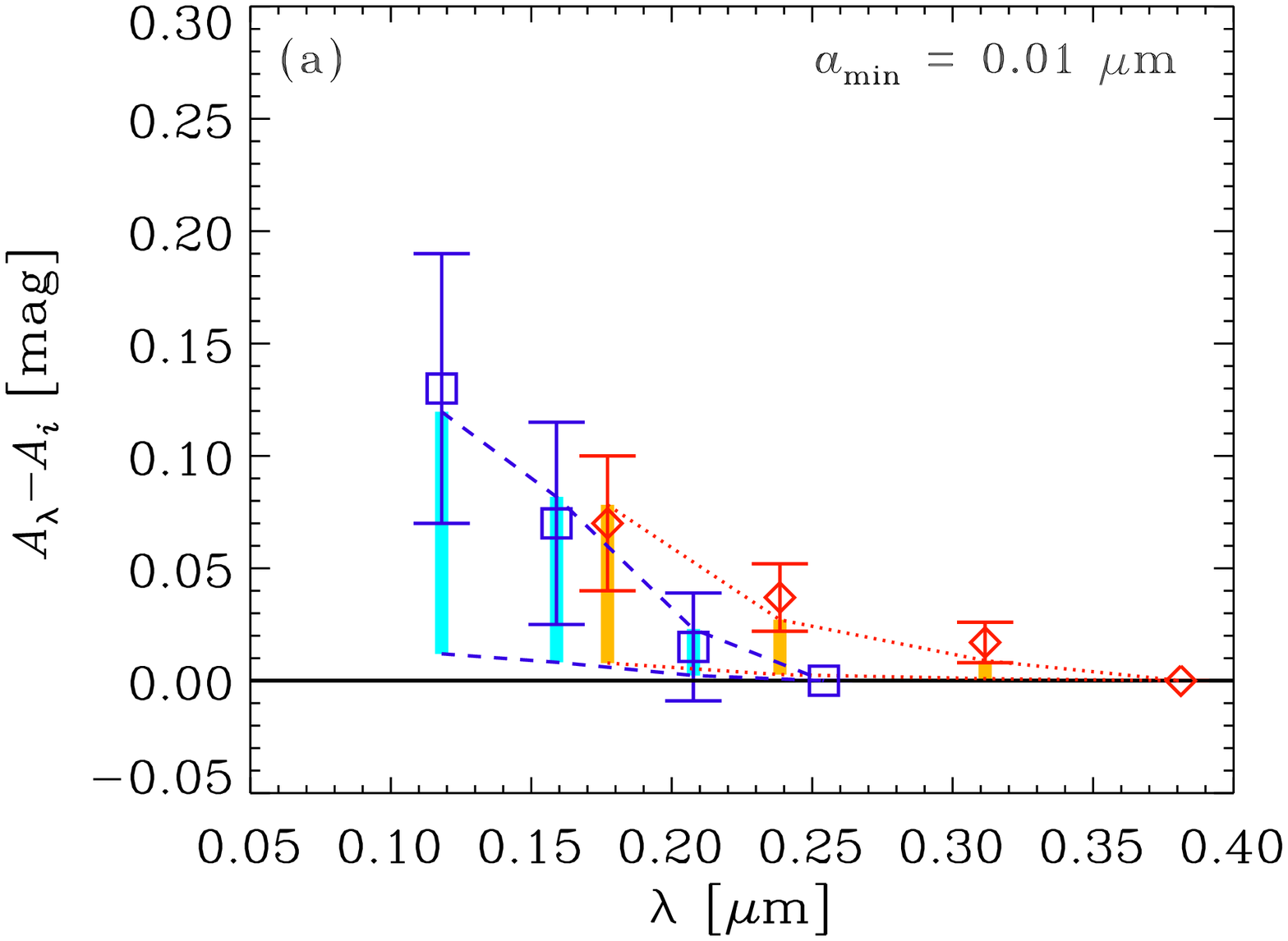}
\includegraphics[width=0.45\textwidth]{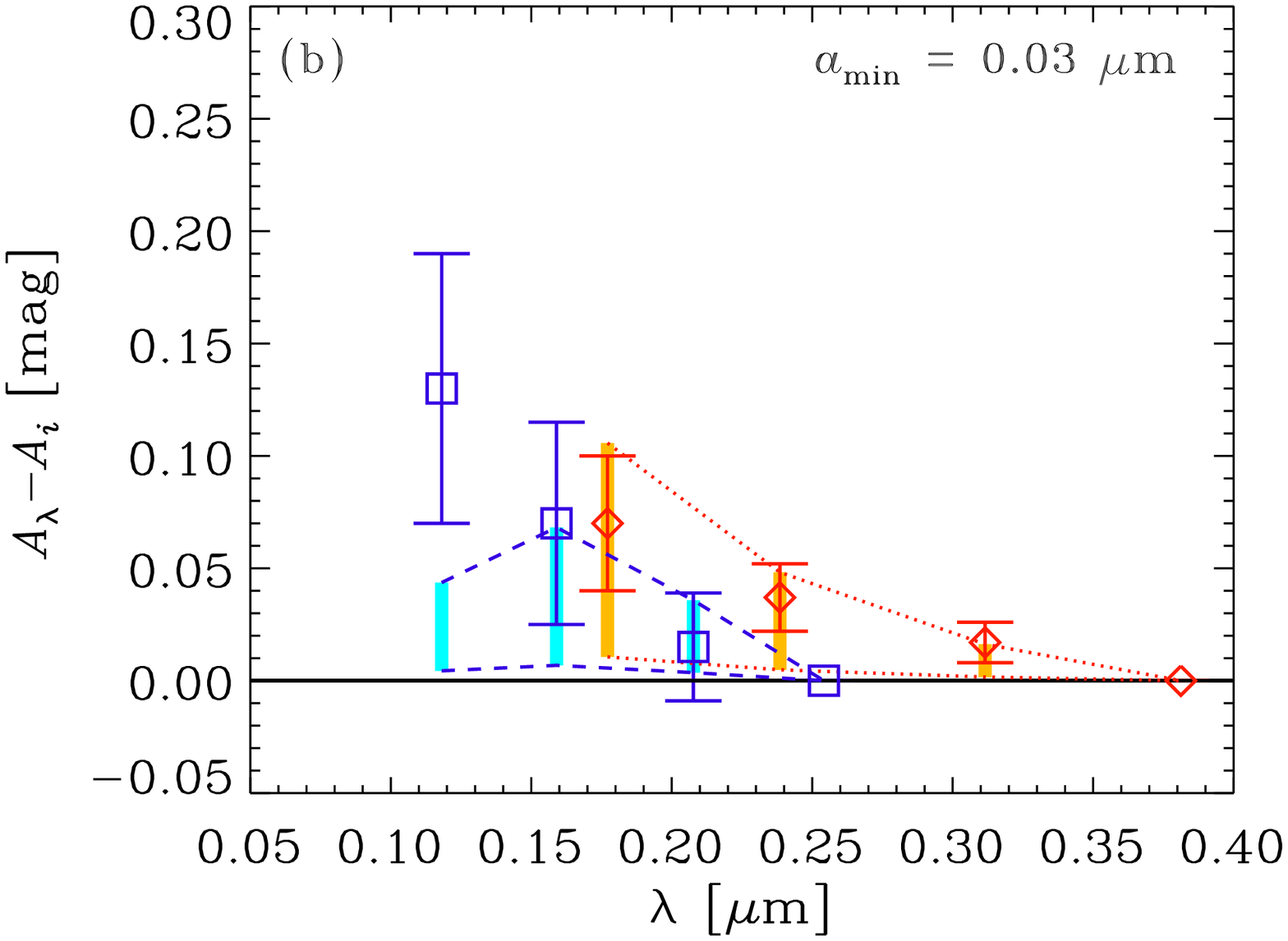}
\includegraphics[width=0.45\textwidth]{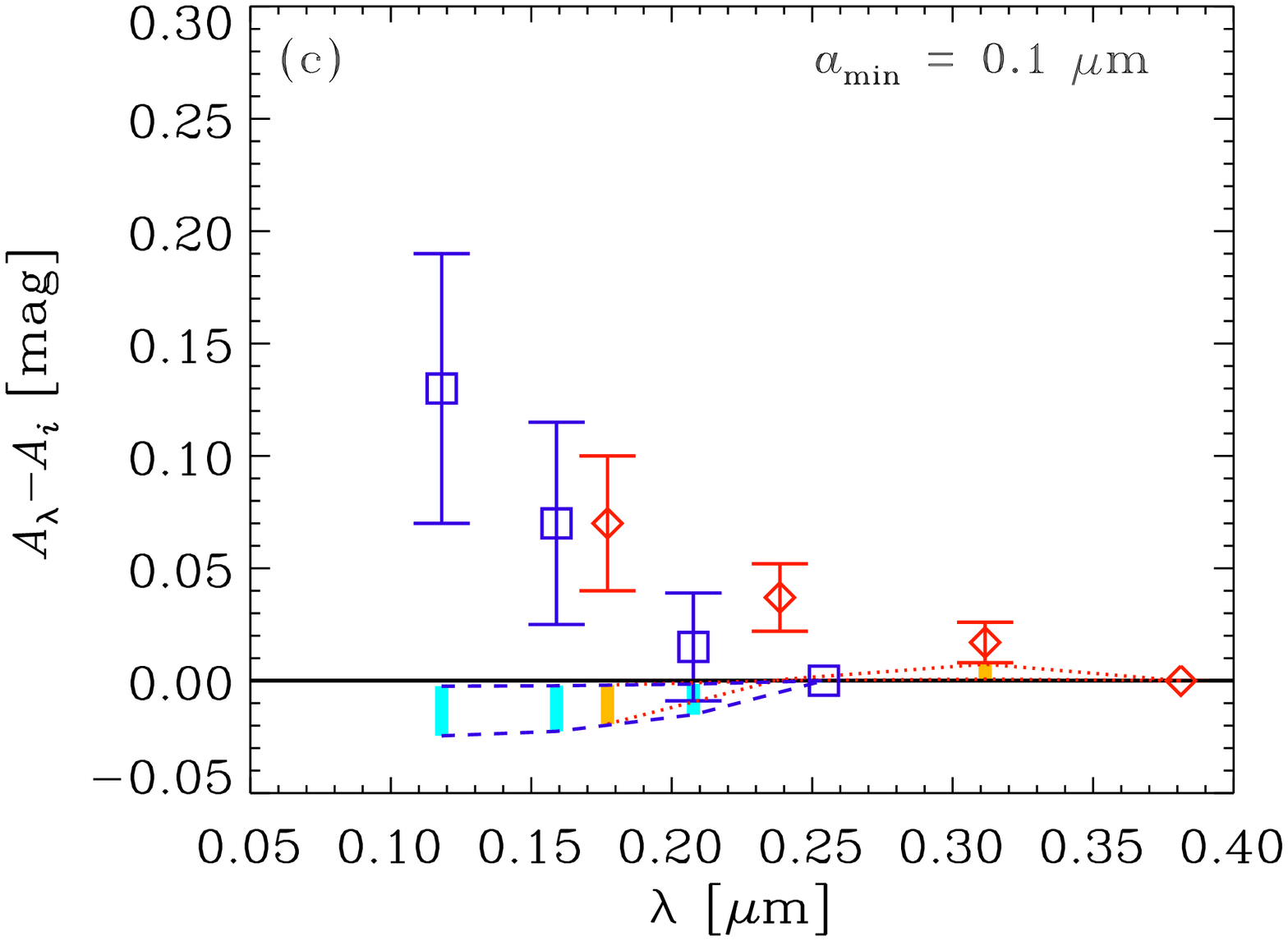}
\caption{{Same as Fig.\ \ref{fig:reddening_sil}
but including the effect of grain size distribution
(equation \ref{eq:MRN} with $a_\mathrm{max}=0.25~\mu$m)
for silicate.
The minimum grain radius is varied as
(a) $a_\mathrm{min}=0.01~\mu$m,
(b) $a_\mathrm{min}=0.03~\mu$m, and
(c) $a_\mathrm{min}=0.1~\mu$m.}
\label{fig:reddening_sizedist_sil}}
\end{figure}

{In Fig.\ \ref{fig:reddening_sizedist_sil}, we show the reddening curve
for silicate with the above grain size distribution. Compared with
Fig.\ \ref{fig:reddening_sil}, the steepness of the reddening curve is moderate
for $a_\mathrm{min}=0.01$ and 0.03 $\mu$m because of the contamination
of large grains. Nevertheless, the theoretical reddening
curves are broadly consistent with the observations if
$a_\mathrm{min}\leq 0.03~\mu$m.
As shown in Fig.\ \ref{fig:reddening_sil}, silicate grains with
$a\sim 0.01~\mu$m have too small extinction, so the major
contribution comes from grains with $a\sim 0.03~\mu$m for
even if $a_\mathrm{min}=0.01~\mu$m.
The reddening curves with $a_\mathrm{min}=0.1~\mu$m is
too flat to explain the observed curves. Therefore, we strengthen the
conclusion that, if the main composition is silicate,
small grains with $a\sim 0.03~\mu$m is necessary
to explain the observed reddening curves in Mg \textsc{ii} absorbers.}

\begin{figure}
\includegraphics[width=0.45\textwidth]{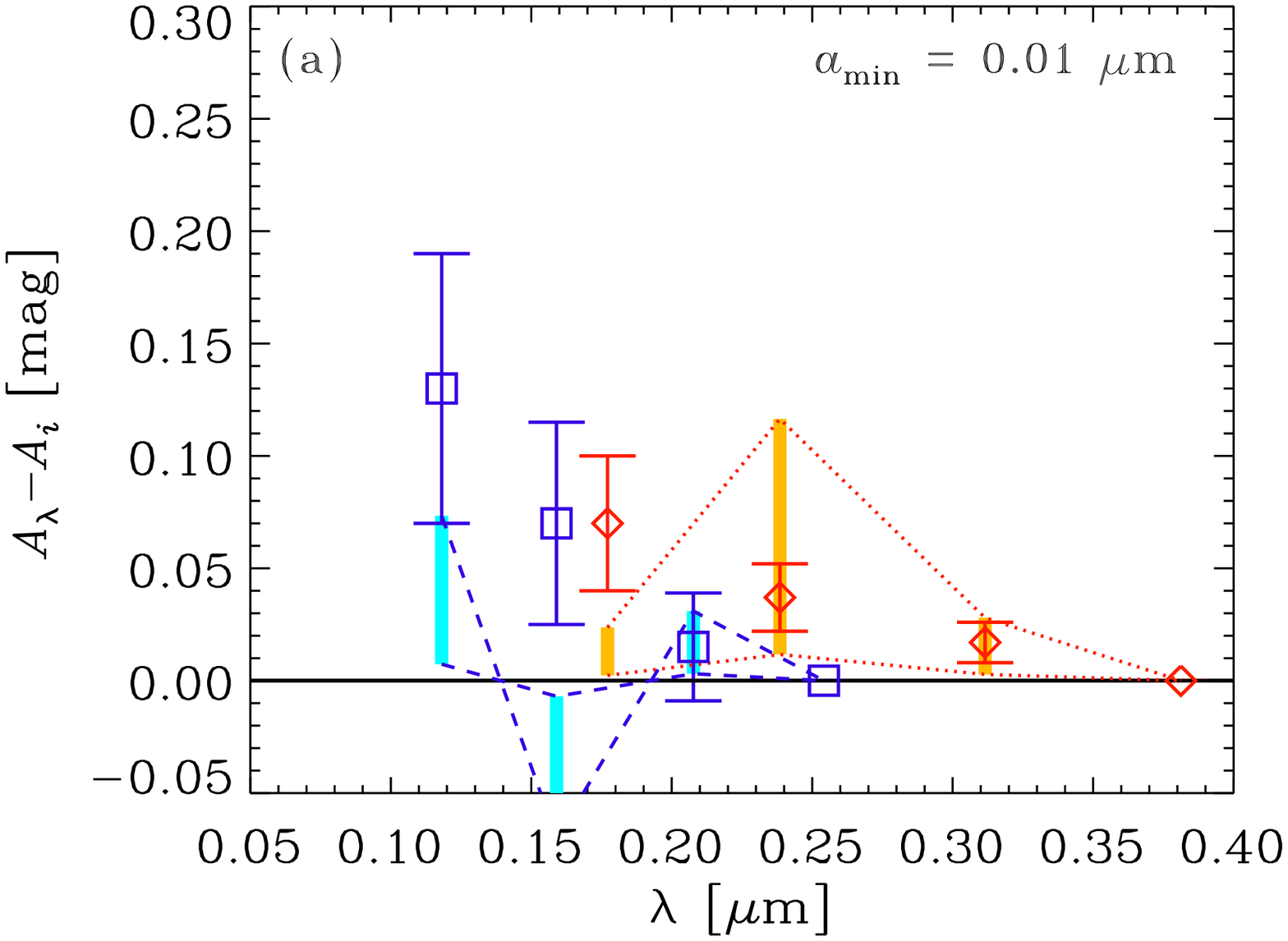}
\includegraphics[width=0.45\textwidth]{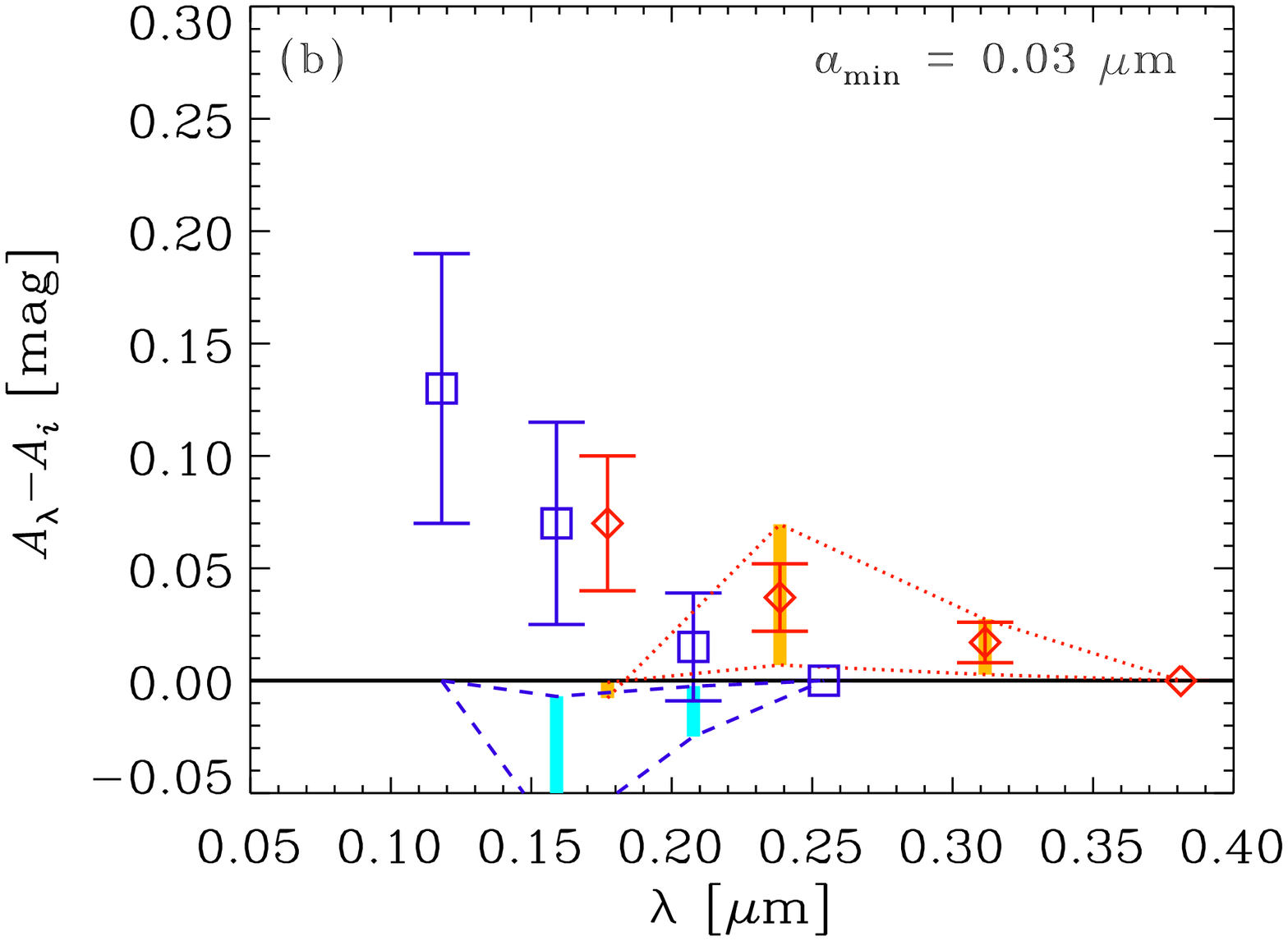}
\includegraphics[width=0.45\textwidth]{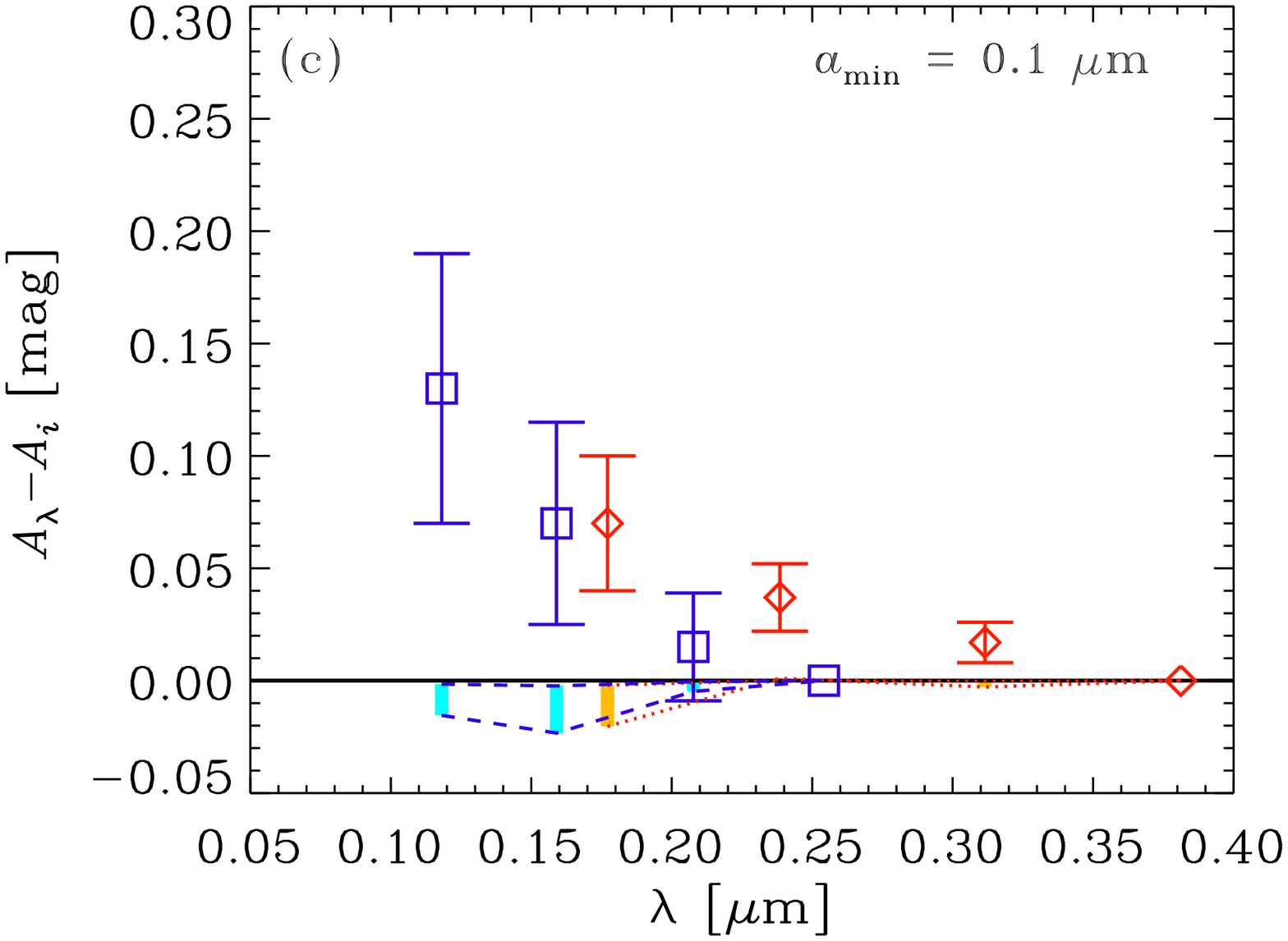}
\caption{{Same as Fig.\ \ref{fig:reddening_sizedist_sil} but
for graphite.}
\label{fig:reddening_sizedist_gra}}
\end{figure}

{In Fig.\ \ref{fig:reddening_sizedist_gra}, we show the results
for graphite. Because of the contamination of large grains, the
bump feature is moderate compared with Fig.\ \ref{fig:reddening_gra};
however, the non-monotonic behavior affected by the bump still remains
if $a_\mathrm{min}\lesssim 0.03~\mu$m. If the minimum grain size
is as large as $\sim 0.1~\mu$m, the reddening curve is too flat to explain the
observations. Therefore, graphite is still not favored even if we consider
the effect of grain size distribution.}

\begin{figure}
\includegraphics[width=0.45\textwidth]{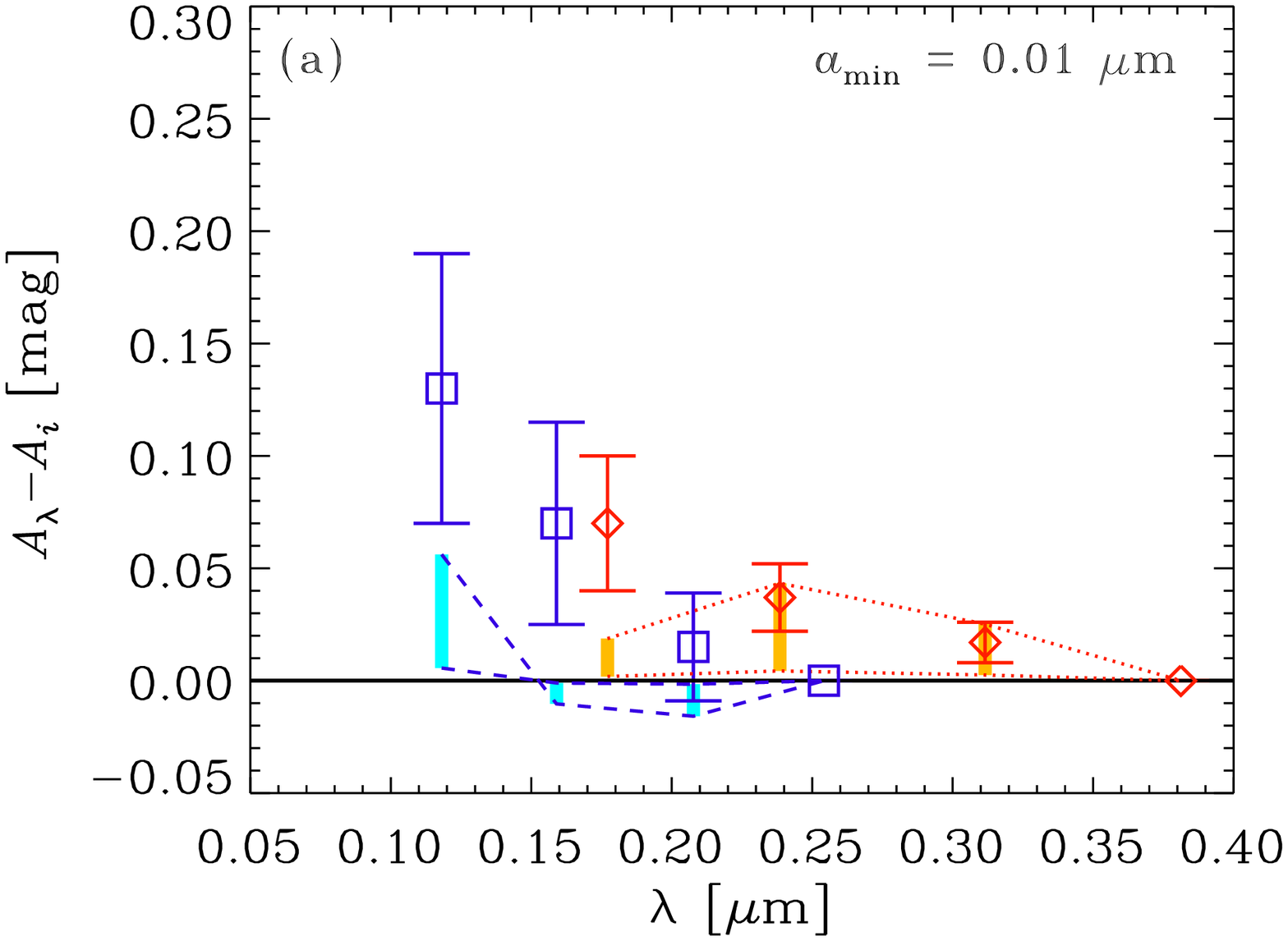}
\includegraphics[width=0.45\textwidth]{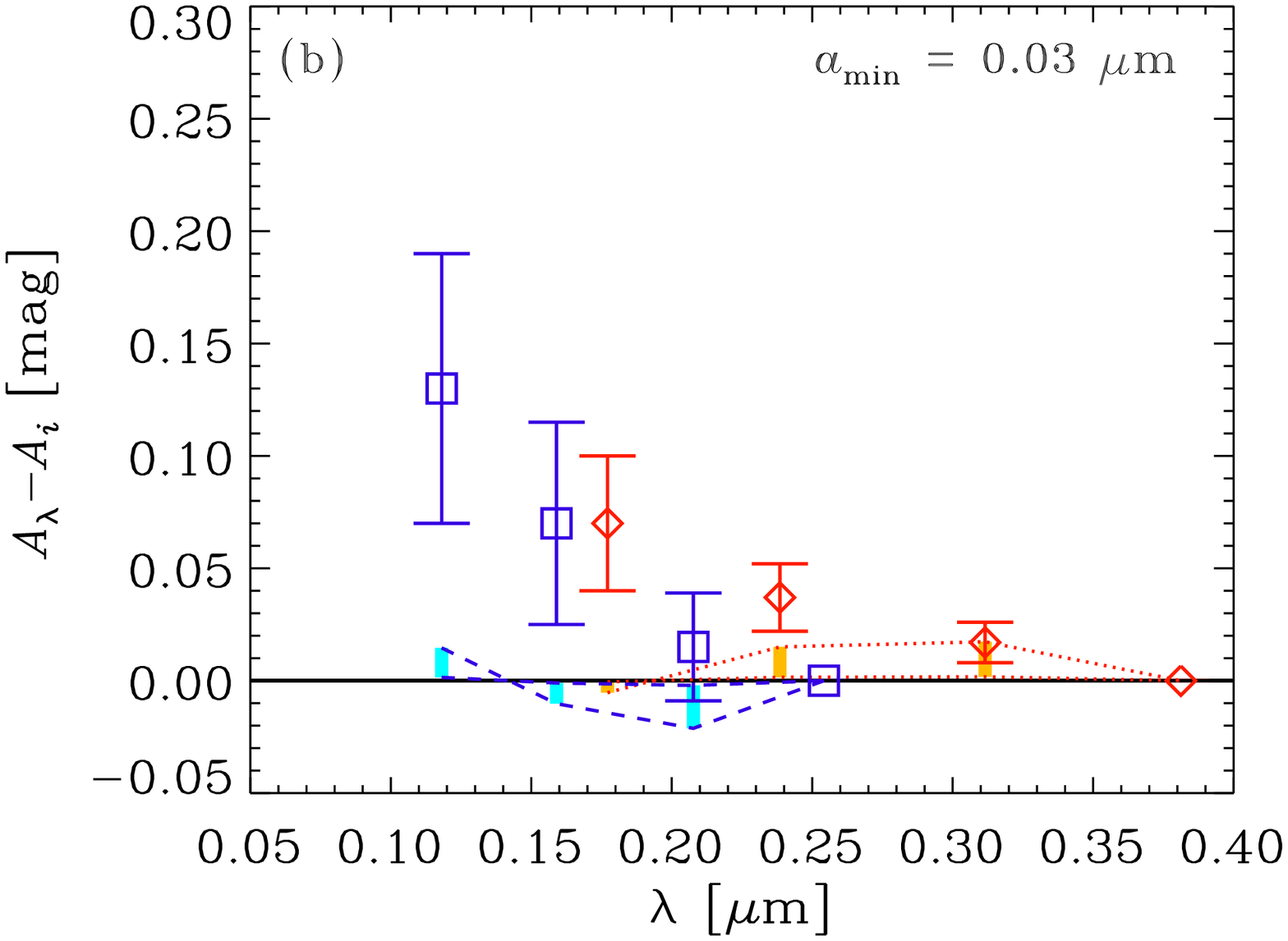}
\includegraphics[width=0.45\textwidth]{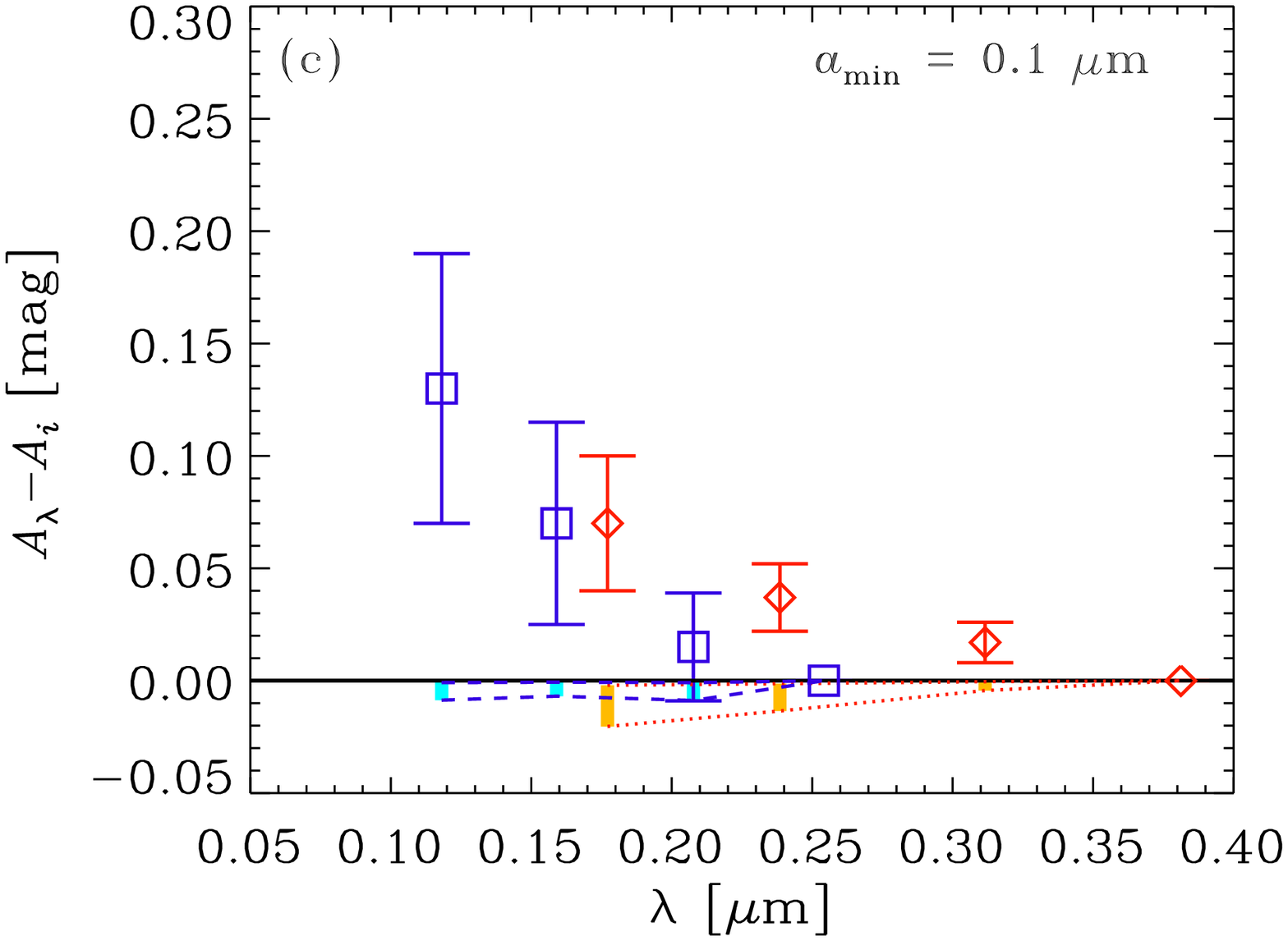}
\caption{{Same as Fig.\ \ref{fig:reddening_sizedist_sil} but
for amorphous carbon.}
\label{fig:reddening_sizedist_AC}}
\end{figure}

{In Fig.\ \ref{fig:reddening_sizedist_AC}, we show the reddening
curves for amorphous carbon.  As expected, the strong bump is
effectively eliminated. However, the calculated reddening curves are
too flat to explain the observations
even with $a_\mathrm{min}=0.01~\mu$m because of the contamination of
large grains. Therefore, if the major component
of dust is amorphous carbon, it is required that most grains have
$a\lesssim 0.03~\mu$m.}

{The conclusions derived with a single grain radius
are not altered as long as the reddening curve is concerned:
if the main composition is silicate, small grains with
$a\lesssim 0.03~\mu$m are required
(i.e., $a_\mathrm{min}\lesssim 0.03~\mu$m).
If the main composition is carbonaceous
dust, the grain size distribution should be biased to small sizes
($a\lesssim 0.03~\mu$m).}

\begin{figure}
\begin{center}
\includegraphics[width=0.4\textwidth]{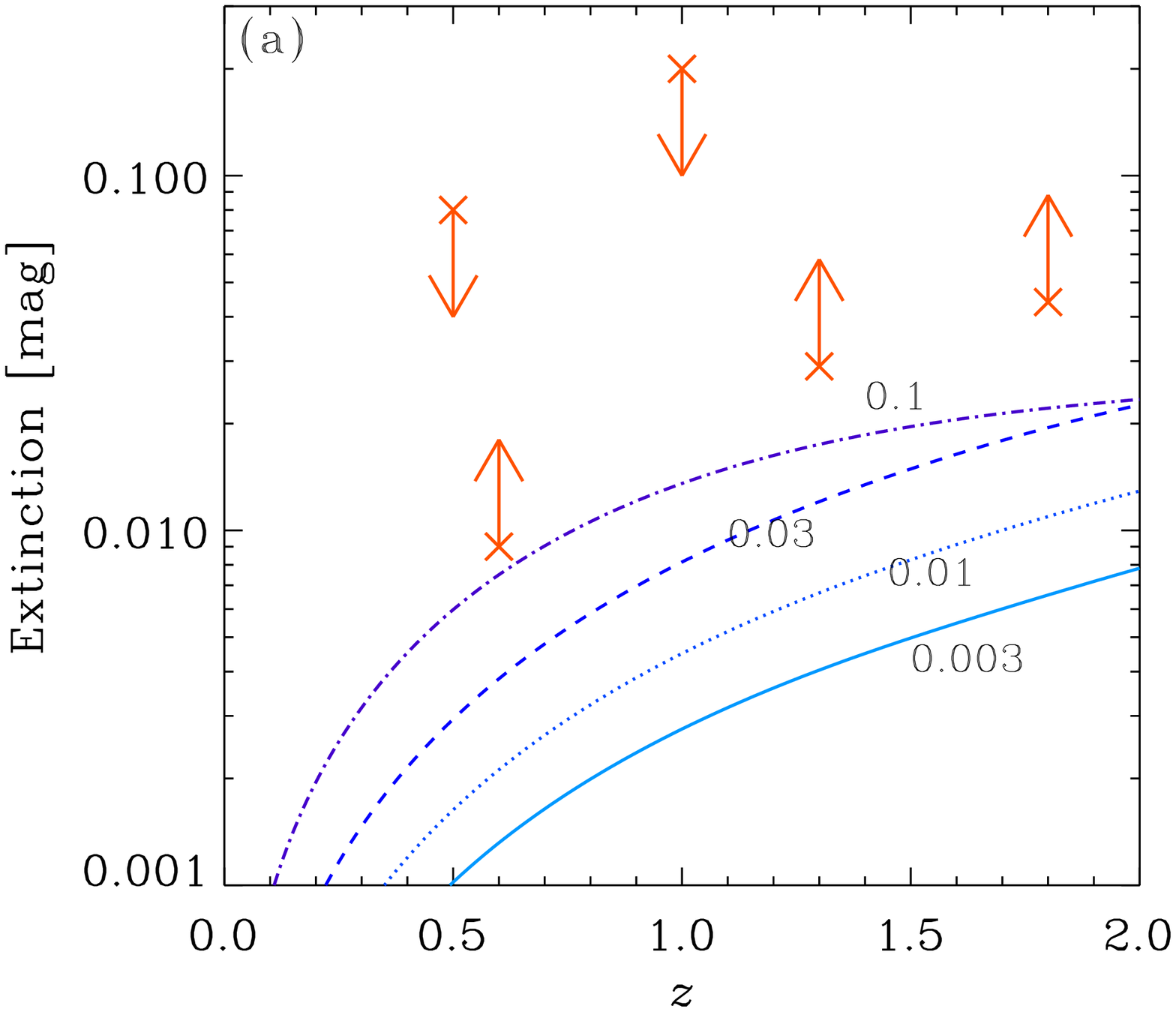}
\includegraphics[width=0.4\textwidth]{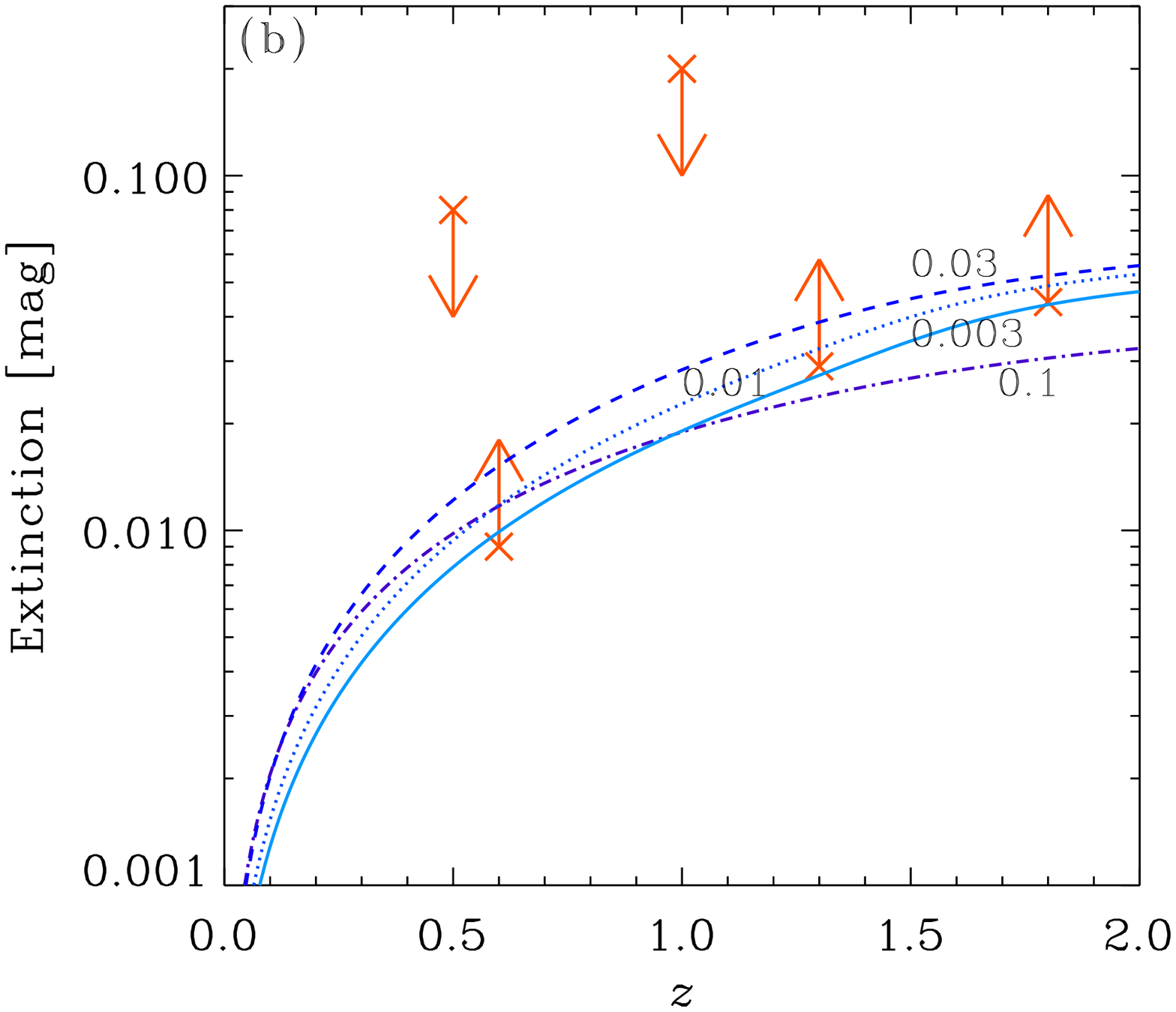}
\includegraphics[width=0.4\textwidth]{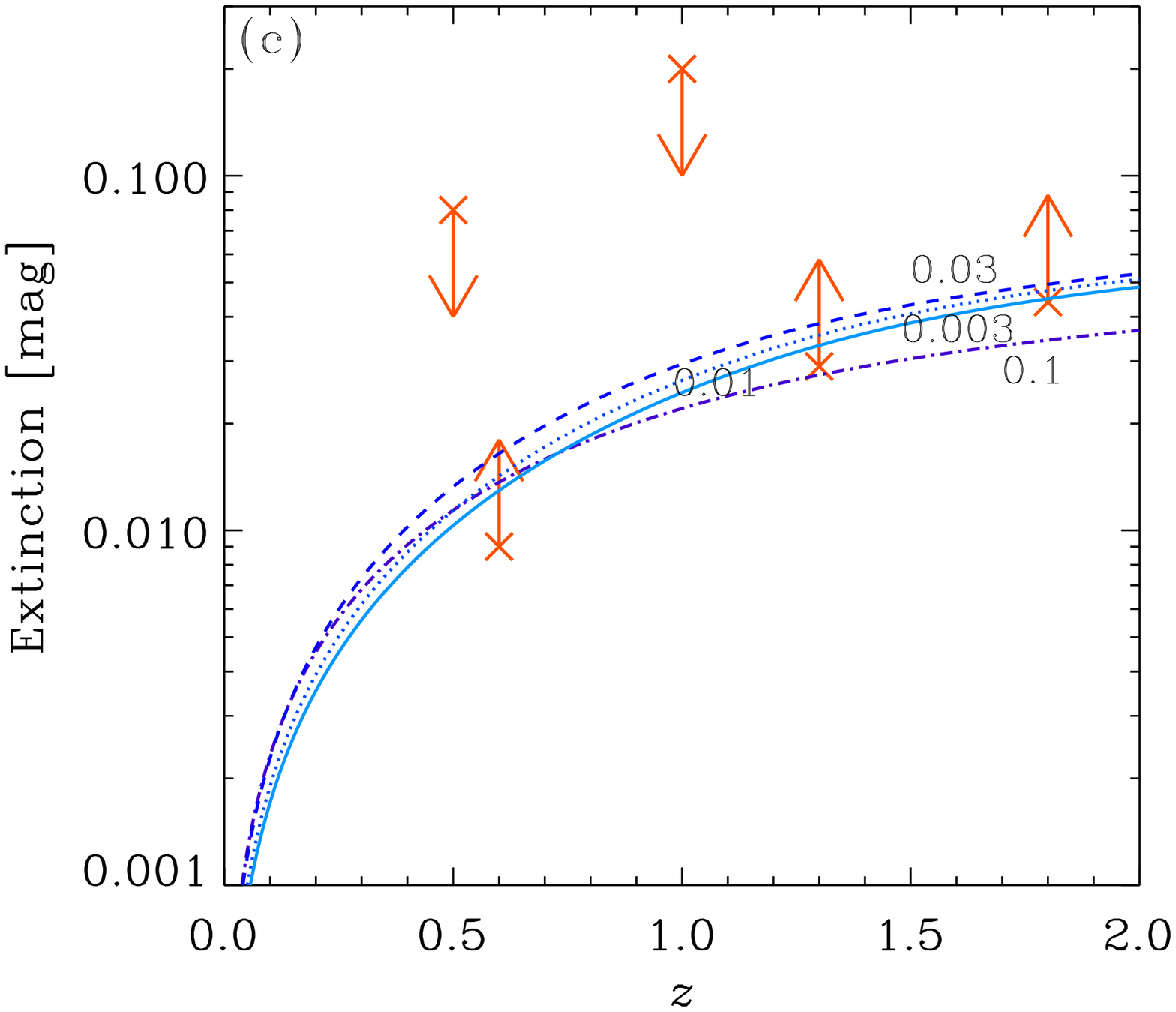}
\end{center}
\caption{{Same as Fig.\ \ref{fig:tau_size} but including the
effect of grain size distribution.
The solid, dotted, dashed, and
dot-dashed lines show the results for
$a_\mathrm{min}=0.003$, 0.01, 0.03, and 0.1 $\mu$m, respectively.}
\label{fig:tau_sizedist_size}}
\end{figure}

{In Fig.\ \ref{fig:tau_sizedist_size}, we also plot the cosmic extinction with
the effect of grain size distribution. We adopt $\eta =10^{-3}$.
Compared with Fig.\ \ref{fig:tau_size},
the variation of cosmic extinction among various $a_\mathrm{min}$
is smaller. For silicate with $a_\mathrm{min}=0.03~\mu$m, which produces a
consistent reddening curve with observations (Fig.\ \ref{fig:reddening_sizedist_sil}),
the theoretical cosmic extinction is 3 times smaller than the
observational lower limits. This discrepancy (3 times) is smaller than
the discrepancy seen in the single-size case with $a=0.03~\mu$m
(10 times; Fig.\ \ref{fig:tau_size}a). This means that, if we take the grain size
distribution into account, we find a solution that explains both the reddening
curves and the cosmic extinction with $\eta\gtrsim 3\times 10^{-3}$. 
For graphite and amorphous carbon,
the calculated cosmic extinction is consistent with the observational constraints.
}

{In summary, the conclusions under the single-radius treatment
above do not change significantly even if we consider the grain
size distribution: (i) For silicate,
inclusion of small $a_\mathrm{min}\lesssim 0.03~\mu$m is favored,
and a larger dust abundance ($\eta\gtrsim 3\times 10^{-3}$)
than suggested by observations is
required. (ii) For carbonaceous dust (amorphous carbon),
small grains with $a\sim 0.03~\mu$m should be dominant.}

\subsection{Comparison with other theoretical studies}

Recently, there have been some cosmological simulations and
semi-analytic models that calculate dust formation in galaxies
and subsequent dust transport
into galaxy halos or intergalactic space
\citep{McKinnon:2016aa,McKinnon:2017aa,Popping:2017aa}.
\citet{McKinnon:2016aa} showed that the radial profile of dust
surface density in the galaxy halo can be roughly
consistent with M10's observational data, depending on the stellar
feedback model adopted \citep[see also][]{Aoyama:2018aa}.
\citet{Popping:2017aa}, using their semi-analytic model, showed that
the galaxy halo component contains a comparable amount of dust to
the ISM component.
\citet{Zu:2011aa} used a cosmological hydrodynamics simulation
and reproduced the
radial distribution of dust in galaxy halos by assuming
that dust traces the metals. They also showed the
dust abundance in galaxy halos is sensitive to the stellar feedback model.
Therefore, it is still a challenge for theoretical
models to reproduce such a large dust abundance in galaxy halos as
observed by M10.

Hydrodynamic simulations with dust enrichment by
\citet{Aoyama:2017aa,Aoyama:2018aa}
and \citet{Hou:2017aa} treat information on grain size distribution
represented by the abundance of large and small grains.
\citet{Hou:2017aa} {and \citet{Aoyama:2018aa}} show that
the grains in galaxy halos are biased to large ($a>0.03~\mu$m) grains.
In contrast, as shown above, the reddening curves of
Mg \textsc{ii} absorbers at $z\sim 1$ and 2 indicate
the dominance of small ($a\lesssim 0.03~\mu$m) grains.
Thus, there is a tension in the grain size between the above simulations and
the observed reddening curves. However, the above two simulations did not
include dust transport by radiation pressure, which could be important as
a mechanism of outflow \citep{Murray:2011aa}.
\citet{Ferrara:1991aa} and \citet{Bianchi:2005aa} show
that grains with $a\sim 0.03$--0.3~$\mu$m can be dynamically decoupled
from gas and transported to
the galaxy halo, but that the radiation pressure has no reason of preferentially
transporting grains with $a\lesssim 0.03~\mu$m.

It may be possible that small grains are produced \textit{in situ}
in Mg \textsc{ii} absorbers.
Some authors argue that Mg \textsc{ii} absorbers are associated with outflow
induced by active star formation \citep{Norman:1996aa,Bond:2001aa}.
Because of a high velocity of outflow, shock or high-velocity turbulence
could be induced in the wind. Both shock \citep{Jones:1996aa} and turbulence
\citep{Yan:2004aa,Hirashita:2009ab}
are considered as mechanisms of inducing grain shattering. Therefore,
grains may be shattered in the outflows.
It is suggested that Mg \textsc{ii} absorbers has a spatial scale 
as small as 30 pc \citep{Lan:2017aa}.
Since the above simulations
are not capable of resolving such a small-scale structure, shattering in
outflowing gas was not successfully treated. In the future,
it may be desirable to investigate the possibility of shattering in
outflows using a high-resolution simulations that are capable of
resolving the internal structures
in the outflowing gas.

\section{Conclusion}\label{sec:conclusion}

We investigate the abundance and properties of
the dust in galaxy halos using available observations in the literature.
There are two major sets of data: reddening curves and cosmic extinction.
The reddening curves are derived by assuming that Mg \textsc{ii}
absorbers trace the medium in galaxy halos. From the reddening curves,
we show that the typical grain radius is $a\sim 0.03~\mu$m for silicate.
Graphite is not consistent with the observed reddening curve because
of its strong 2175 \AA\ bump. Using amorphous carbon improves
the fit to the reddening. For amorphous carbon, $a\lesssim 0.03~\mu$m
is supported by the reddening curves.

The observational constraints on the cosmic extinction suggest that
the grain radius be $a\lesssim 0.1~\mu$m for carbonaceous dust
if the halo dust abundance relative to the stellar mass
is $\eta =10^{-3}$ as suggested by M10.
For silicate, if $\eta =10^{-3}$, the cosmic extinction is significantly
underestimated and only $a\sim 0.1~\mu$m is marginally consistent with
the observational lower limits. In other words, silicate
favors a higher dust abundance in galaxy halos such as
$\eta\gtrsim 10^{-2}$ if we adopt $a\sim 0.03~\mu$m suggested
by the reddening curves.
{The constraint on $\eta$ becomes milder ($\eta\gtrsim 3\times 10^{-3}$)
if we take a grain size distribution into account, but the value of $\eta$ is
still larger than the fiducial value.}
Thus, if the main dust composition in galaxy halos is silicate,
there is a tension in the grain size between the
constraint from the reddening and that from the cosmic extinction.
For carbonaceous dust (amorphous carbon), those two constraints
indicate $a\lesssim 0.03~\mu$m.
{The requirement of small grains with $a\lesssim 0.03~\mu$m is
robustly concluded even if we consider grain size distribution.}

The favored grain radii ($a\lesssim 0.03~\mu$m) in galaxy halos are not
in line with numerical simulations by \citet{Hou:2017aa} and
{\citet{Aoyama:2018aa}}.
In their scenario, the halo dust is dominated by large
($a > 0.03~\mu$m) grains originating from stars because the dust tends to
be ejected by stellar feedback before being significantly processed in the ISM
by shattering and accretion. Although the above simulations do not include the
effect of radiation pressure, there is no physical reason that small
grains are selectively transported by radiation pressure.
We point out a possibility of shattering in the outflow driven by
stellar feedback, but a high resolution simulation is necessary to
examine whether or not this happens.

\section*{Acknowledgment}

We are grateful to S. Aoyama, K. Nagamine, K.-C. Hou, S. Bianchi,
and the anonymous referee for their useful discussions and comments.
HH thanks the Ministry of Science and Technology for support through grant
MOST 105-2112-M-001-027-MY3 and MOST 107-2923-M-001-003-MY3.

\appendix

\section{Comoving mass densities}
\label{app:comoving}

The comoving stellar mass density, $\rho_\mathrm{star}$, is
estimated using the stellar mass function as
\begin{eqnarray}
\rho_\mathrm{star}=\int M_\mathrm{star}\Phi (M_\mathrm{star})\,
\mathrm{d}M_\mathrm{star},\label{eq:rho_star}
\end{eqnarray}
while the comoving halo dust mass density, $\rho_\mathrm{d,halo}$,
is estimated as
\begin{eqnarray}
\rho_\mathrm{d,halo}=\int M_\mathrm{d,halo}\Phi (M_\mathrm{star})\,
\mathrm{d}M_\mathrm{star}.\label{eq:rho_d}
\end{eqnarray}
We also evaluate the comoving number density of galaxies:
\begin{eqnarray}
n_\mathrm{g}=\int\Phi (M_\mathrm{star})\,
\mathrm{d}M_\mathrm{star}.\label{eq:number}
\end{eqnarray}

Now we define the averaged halo dust-to-stellar mass ratio,
$\langle\eta\rangle$ as
\begin{eqnarray}
\langle\eta\rangle & \hspace{-2mm} = & \hspace{-2mm}
\int M_\mathrm{d,halo}\Phi (M_\mathrm{star})\,
\mathrm{d}M_\mathrm{star}\left/\int M_\mathrm{star}\Phi (M_\mathrm{star})\,
\mathrm{d}M_\mathrm{star}\right.\nonumber\\
& \hspace{-2mm} = & \hspace{-2mm}
\int\eta M_\mathrm{star}\Phi (M_\mathrm{star})\,
\mathrm{d}M_\mathrm{star}\left/
\int M_\mathrm{star}\Phi (M_\mathrm{star})\,
\mathrm{d}M_\mathrm{star}\right. .\label{eq:eta_av_def}
\end{eqnarray}
Using equations (\ref{eq:rho_star}), (\ref{eq:rho_d}), and
(\ref{eq:eta_av_def}), we obtain equation (\ref{eq:eta_av}).

We also define the average halo dust mass as
\begin{eqnarray}
\langle M_\mathrm{d,halo}\rangle =\int M_\mathrm{d,halo}\Phi (M_\mathrm{star})\,
\mathrm{d}M_\mathrm{star}\left/\int\Phi (M_\mathrm{star})\,
\mathrm{d}M_\mathrm{star}\right. .\label{eq:Md_av_def}
\end{eqnarray}
Using equations (\ref{eq:rho_d}), (\ref{eq:number}) and
(\ref{eq:Md_av_def}), we obtain
\begin{eqnarray}
\rho_\mathrm{d,halo}=n_\mathrm{g}\langle M_\mathrm{d,halo}\rangle .
\label{eq:rho_dhalo}
\end{eqnarray}

\bibliographystyle{elsarticle-harv} 
\bibliography{hirashita1}

\begin{thebibliography}{53}
\expandafter\ifx\csname natexlab\endcsname\relax\def\natexlab#1{#1}\fi
\expandafter\ifx\csname url\endcsname\relax
  \def\url#1{\texttt{#1}}\fi
\expandafter\ifx\csname urlprefix\endcsname\relax\def\urlprefix{URL }\fi

\bibitem[{{Aguirre}(1999)}]{Aguirre:1999aa}
{Aguirre}, A., 1999. {Intergalactic Dust and Observations of Type IA
  Supernovae}. \apj 525, 583--593.

\bibitem[{{Aoyama} et~al.(2018){Aoyama}, {Hou}, {Hirashita}, {Nagamine}, and
  {Shimizu}}]{Aoyama:2018aa}
{Aoyama}, S., {Hou}, K.-C., {Hirashita}, H., {Nagamine}, K., {Shimizu}, I.,
  2018. {Cosmological simulation with dust formation and destruction}. ArXiv
  e-prints.

\bibitem[{{Aoyama} et~al.(2017){Aoyama}, {Hou}, {Shimizu}, {Hirashita},
  {Todoroki}, {Choi}, and {Nagamine}}]{Aoyama:2017aa}
{Aoyama}, S., {Hou}, K.-C., {Shimizu}, I., {Hirashita}, H., {Todoroki}, K.,
  {Choi}, J.-H., {Nagamine}, K., 2017. {Galaxy simulation with dust formation
  and destruction}. \mnras 466, 105--121.

\bibitem[{{Asano} et~al.(2013){Asano}, {Takeuchi}, {Hirashita}, and
  {Nozawa}}]{Asano:2013aa}
{Asano}, R.~S., {Takeuchi}, T.~T., {Hirashita}, H., {Nozawa}, T., 2013. {What
  determines the grain size distribution in galaxies?} \mnras 432, 637--652.

\bibitem[{{Avgoustidis} et~al.(2009){Avgoustidis}, {Verde}, and
  {Jimenez}}]{Avgoustidis:2009aa}
{Avgoustidis}, A., {Verde}, L., {Jimenez}, R., 2009. {Consistency among
  distance measurements: transparency, BAO scale and accelerated expansion}.
  \jcap 6, 012.

\bibitem[{{Bekki}(2015)}]{Bekki:2015ab}
{Bekki}, K., 2015. {Dust-regulated galaxy formation and evolution: a new
  chemodynamical model with live dust particles}. \mnras 449, 1625--1649.

\bibitem[{{Bianchi} and {Ferrara}(2005)}]{Bianchi:2005aa}
{Bianchi}, S., {Ferrara}, A., 2005. {Intergalactic medium metal enrichment
  through dust sputtering}. \mnras 358, 379--396.

\bibitem[{{Bohren} and {Huffman}(1983)}]{Bohren:1983aa}
{Bohren}, C.~F., {Huffman}, D.~R., 1983. Absorption and Scattering of Light by
  Small Particles. Wiley.

\bibitem[{{Bond} et~al.(2001){Bond}, {Churchill}, {Charlton}, and
  {Vogt}}]{Bond:2001aa}
{Bond}, N.~A., {Churchill}, C.~W., {Charlton}, J.~C., {Vogt}, S.~S., 2001.
  {High-Redshift Superwinds as the Source of the Strongest Mg II Absorbers: A
  Feasibility Analysis}. \apj 562, 641--648.

\bibitem[{{Bouch{\'e}} et~al.(2007){Bouch{\'e}}, {Murphy}, {P{\'e}roux},
  {Davies}, {Eisenhauer}, {F{\"o}rster Schreiber}, and
  {Tacconi}}]{Bouche:2007aa}
{Bouch{\'e}}, N., {Murphy}, M.~T., {P{\'e}roux}, C., {Davies}, R.,
  {Eisenhauer}, F., {F{\"o}rster Schreiber}, N.~M., {Tacconi}, L., 2007. {The
  SINFONI Mg II Program for Line Emitters (SIMPLE): Discovering Starbursts near
  QSO Sight Lines}. \apjl 669, L5--L8.

\bibitem[{{Draine} and {Lee}(1984)}]{Draine:1984aa}
{Draine}, B.~T., {Lee}, H.~M., 1984. {Optical properties of interstellar
  graphite and silicate grains}. \apj 285, 89--108.

\bibitem[{{Evans}(1994)}]{Evans:1994aa}
{Evans}, A., 1994. The Dusty Universe. Wiley.

\bibitem[{{Ferrara} et~al.(1991){Ferrara}, {Ferrini}, {Barsella}, and
  {Franco}}]{Ferrara:1991aa}
{Ferrara}, A., {Ferrini}, F., {Barsella}, B., {Franco}, J., 1991. {Evolution of
  dust grains through a hot gaseous halo}. \apj 381, 137--146.

\bibitem[{{Hildebrand}(1983)}]{Hildebrand:1983aa}
{Hildebrand}, R.~H., 1983. {The Determination of Cloud Masses and Dust
  Characteristics from Submillimetre Thermal Emission}. \qjras 24, 267.

\bibitem[{{Hirashita} and {Yan}(2009)}]{Hirashita:2009ab}
{Hirashita}, H., {Yan}, H., 2009. {Shattering and coagulation of dust grains in
  interstellar turbulence}. \mnras 394, 1061--1074.

\bibitem[{{Hou} et~al.(2016){Hou}, {Hirashita}, and
  {Micha{\l}owski}}]{Hou:2016aa}
{Hou}, K.-C., {Hirashita}, H., {Micha{\l}owski}, M.~J., 2016. {Dust evolution
  processes constrained by extinction curves in nearby galaxies}. \pasj 68, 94.

\bibitem[{{Hou} et~al.(2017){Hou}, {Hirashita}, {Nagamine}, {Aoyama}, and
  {Shimizu}}]{Hou:2017aa}
{Hou}, K.-C., {Hirashita}, H., {Nagamine}, K., {Aoyama}, S., {Shimizu}, I.,
  2017. {Evolution of dust extinction curves in galaxy simulation}. \mnras 469,
  870--885.

\bibitem[{{Inoue} and {Kamaya}(2003)}]{Inoue:2003ab}
{Inoue}, A.~K., {Kamaya}, H., 2003. {Constraint on intergalactic dust from
  thermal history of intergalactic medium}. \mnras 341, L7--L11.

\bibitem[{{Inoue} and {Kamaya}(2004)}]{Inoue:2004aa}
{Inoue}, A.~K., {Kamaya}, H., 2004. {Amount of intergalactic dust: constraints
  from distant supernovae and the thermal history of the intergalactic medium}.
  \mnras 350, 729--744.

\bibitem[{{Inoue} and {Kamaya}(2010)}]{Inoue:2010aa}
{Inoue}, A.~K., {Kamaya}, H., 2010. {Intergalactic dust and its photoelectric
  heating}. Earth, Planets, and Space 62, 69--79.

\bibitem[{{Jenkins}(2009)}]{Jenkins:2009aa}
{Jenkins}, E.~B., 2009. {A Unified Representation of Gas-Phase Element
  Depletions in the Interstellar Medium}. \apj 700, 1299--1348.

\bibitem[{{Jones} et~al.(2013){Jones}, {Fanciullo}, {K{\"o}hler}, {Verstraete},
  {Guillet}, {Bocchio}, and {Ysard}}]{Jones:2013aa}
{Jones}, A.~P., {Fanciullo}, L., {K{\"o}hler}, M., {Verstraete}, L., {Guillet},
  V., {Bocchio}, M., {Ysard}, N., 2013. {The evolution of amorphous
  hydrocarbons in the ISM: dust modelling from a new vantage point}. \aap 558,
  A62.

\bibitem[{{Jones} et~al.(1996){Jones}, {Tielens}, and
  {Hollenbach}}]{Jones:1996aa}
{Jones}, A.~P., {Tielens}, A.~G.~G.~M., {Hollenbach}, D.~J., 1996. {Grain
  Shattering in Shocks: The Interstellar Grain Size Distribution}. \apj 469,
  740.

\bibitem[{{Kere{\v s}} et~al.(2005){Kere{\v s}}, {Katz}, {Weinberg}, and
  {Dav{\'e}}}]{Keres:2005aa}
{Kere{\v s}}, D., {Katz}, N., {Weinberg}, D.~H., {Dav{\'e}}, R., 2005. {How do
  galaxies get their gas?} \mnras 363, 2--28.

\bibitem[{{Lan} and {Fukugita}(2017)}]{Lan:2017aa}
{Lan}, T.-W., {Fukugita}, M., 2017. {Mg II Absorbers: Metallicity Evolution and
  Cloud Morphology}. \apj 850, 156.

\bibitem[{{Martin} et~al.(2005){Martin}, {Fanson}, {Schiminovich}, {Morrissey},
  {Friedman}, {Barlow}, {Conrow}, {Grange}, {Jelinsky}, {Milliard}, {Siegmund},
  {Bianchi}, {Byun}, {Donas}, {Forster}, {Heckman}, {Lee}, {Madore}, {Malina},
  {Neff}, {Rich}, {Small}, {Surber}, {Szalay}, {Welsh}, and
  {Wyder}}]{Martin:2005aa}
{Martin}, D.~C., {Fanson}, J., {Schiminovich}, D., {Morrissey}, P., {Friedman},
  P.~G., {Barlow}, T.~A., {Conrow}, T., {Grange}, R., {Jelinsky}, P.~N.,
  {Milliard}, B., {Siegmund}, O.~H.~W., {Bianchi}, L., {Byun}, Y.-I., {Donas},
  J., {Forster}, K., {Heckman}, T.~M., {Lee}, Y.-W., {Madore}, B.~F., {Malina},
  R.~F., {Neff}, S.~G., {Rich}, R.~M., {Small}, T., {Surber}, F., {Szalay},
  A.~S., {Welsh}, B., {Wyder}, T.~K., 2005. {The Galaxy Evolution Explorer: A
  Space Ultraviolet Survey Mission}. \apjl 619, L1--L6.

\bibitem[{{Masaki} and {Yoshida}(2012)}]{Masaki:2012aa}
{Masaki}, S., {Yoshida}, N., 2012. {Distribution of dust around galaxies: an
  analytic model}. \mnras 423, L117--L121.

\bibitem[{{Mathis} et~al.(1977){Mathis}, {Rumpl}, and
  {Nordsieck}}]{Mathis:1977aa}
{Mathis}, J.~S., {Rumpl}, W., {Nordsieck}, K.~H., Oct. 1977. {The size
  distribution of interstellar grains}. \apj 217, 425--433.

\bibitem[{{McKinnon} et~al.(2016){McKinnon}, {Torrey}, and
  {Vogelsberger}}]{McKinnon:2016aa}
{McKinnon}, R., {Torrey}, P., {Vogelsberger}, M., 2016. {Dust formation in
  Milky Way-like galaxies}. \mnras 457, 3775--3800.

\bibitem[{{McKinnon} et~al.(2017){McKinnon}, {Torrey}, {Vogelsberger},
  {Hayward}, and {Marinacci}}]{McKinnon:2017aa}
{McKinnon}, R., {Torrey}, P., {Vogelsberger}, M., {Hayward}, C.~C.,
  {Marinacci}, F., 2017. {Simulating the dust content of galaxies: successes
  and failures}. \mnras 468, 1505--1521.

\bibitem[{{M{\'e}nard} and {Chelouche}(2009)}]{Menard:2009aa}
{M{\'e}nard}, B., {Chelouche}, D., 2009. {On the HI content, dust-to-gas ratio
  and nature of MgII absorbers}. \mnras 393, 808--815.

\bibitem[{{M{\'e}nard} and {Fukugita}(2012)}]{Menard:2012aa}
{M{\'e}nard}, B., {Fukugita}, M., 2012. {Cosmic Dust in Mg II Absorbers}. \apj
  754, 116.

\bibitem[{{M{\'e}nard} et~al.(2008){M{\'e}nard}, {Nestor}, {Turnshek},
  {Quider}, {Richards}, {Chelouche}, and {Rao}}]{Menard:2008aa}
{M{\'e}nard}, B., {Nestor}, D., {Turnshek}, D., {Quider}, A., {Richards}, G.,
  {Chelouche}, D., {Rao}, S., 2008. {Lensing, reddening and extinction effects
  of MgII absorbers from z = 0.4 to 2}. \mnras 385, 1053--1066.

\bibitem[{{M{\'e}nard} et~al.(2010){M{\'e}nard}, {Scranton}, {Fukugita}, and
  {Richards}}]{Menard:2010aa}
{M{\'e}nard}, B., {Scranton}, R., {Fukugita}, M., {Richards}, G., 2010.
  {Measuring the galaxy-mass and galaxy-dust correlations through magnification
  and reddening}. \mnras 405, 1025--1039.

\bibitem[{{M{\"o}rtsell} and {Goobar}(2003)}]{Mortsell:2003aa}
{M{\"o}rtsell}, E., {Goobar}, A., 2003. {Constraints on intergalactic dust from
  quasar colours}. \jcap 9, 009.

\bibitem[{{Murray} et~al.(2011){Murray}, {M{\'e}nard}, and
  {Thompson}}]{Murray:2011aa}
{Murray}, N., {M{\'e}nard}, B., {Thompson}, T.~A., 2011. {Radiation Pressure
  from Massive Star Clusters as a Launching Mechanism for Super-galactic
  Winds}. \apj 735, 66.

\bibitem[{{Norman} et~al.(1996){Norman}, {Bowen}, {Heckman}, {Blades}, and
  {Danly}}]{Norman:1996aa}
{Norman}, C.~A., {Bowen}, D.~V., {Heckman}, T., {Blades}, C., {Danly}, L.,
  1996. {Hubble Space Telescope Observations of QSO Absorption Lines Associated
  with Starburst Galaxy Outflows}. \apj 472, 73.

\bibitem[{{Nozawa} et~al.(2015){Nozawa}, {Asano}, {Hirashita}, and
  {Takeuchi}}]{Nozawa:2015aa}
{Nozawa}, T., {Asano}, R.~S., {Hirashita}, H., {Takeuchi}, T.~T., 2015.
  {Evolution of grain size distribution in high-redshift dusty quasars:
  integrating large amounts of dust and unusual extinction curves}. \mnras 447,
  L16--L20.

\bibitem[{{Peek} et~al.(2015){Peek}, {M{\'e}nard}, and
  {Corrales}}]{Peek:2015aa}
{Peek}, J.~E.~G., {M{\'e}nard}, B., {Corrales}, L., 2015. {Dust in the
  Circumgalactic Medium of Low-redshift Galaxies}. \apj 813, 7.

\bibitem[{{Pei}(1992)}]{Pei:1992aa}
{Pei}, Y.~C., 1992. {Interstellar dust from the Milky Way to the Magellanic
  Clouds}. \apj 395, 130--139.

\bibitem[{{Popping} et~al.(2017){Popping}, {Somerville}, and
  {Galametz}}]{Popping:2017aa}
{Popping}, G., {Somerville}, R.~S., {Galametz}, M., 2017. {The dust content of
  galaxies from $z = 0$ to $z = 9$}. \mnras 471, 3152--3185.

\bibitem[{{Steidel} et~al.(1997){Steidel}, {Dickinson}, {Meyer}, {Adelberger},
  and {Sembach}}]{Steidel:1997aa}
{Steidel}, C.~C., {Dickinson}, M., {Meyer}, D.~M., {Adelberger}, K.~L.,
  {Sembach}, K.~R., 1997. {Quasar Absorbing Galaxies at $z\lesssim 1$: Deep
  Imaging and Spectroscopy in the Field of 3C 336}. \apj 480, 568--588.

\bibitem[{{Tomczak} et~al.(2014){Tomczak}, {Quadri}, {Tran}, {Labb{\'e}},
  {Straatman}, {Papovich}, {Glazebrook}, {Allen}, {Brammer}, {Kacprzak},
  {Kawinwanichakij}, {Kelson}, {McCarthy}, {Mehrtens}, {Monson}, {Persson},
  {Spitler}, {Tilvi}, and {van Dokkum}}]{Tomczak:2014aa}
{Tomczak}, A.~R., {Quadri}, R.~F., {Tran}, K.-V.~H., {Labb{\'e}}, I.,
  {Straatman}, C.~M.~S., {Papovich}, C., {Glazebrook}, K., {Allen}, R.,
  {Brammer}, G.~B., {Kacprzak}, G.~G., {Kawinwanichakij}, L., {Kelson}, D.~D.,
  {McCarthy}, P.~J., {Mehrtens}, N., {Monson}, A.~J., {Persson}, S.~E.,
  {Spitler}, L.~R., {Tilvi}, V., {van Dokkum}, P., 2014. {Galaxy Stellar Mass
  Functions from ZFOURGE/CANDELS: An Excess of Low-mass Galaxies since $z = 2$
  and the Rapid Buildup of Quiescent Galaxies}. \apj 783, 85.

\bibitem[{{Tremonti} et~al.(2007){Tremonti}, {Moustakas}, and
  {Diamond-Stanic}}]{Tremonti:2007aa}
{Tremonti}, C.~A., {Moustakas}, J., {Diamond-Stanic}, A.~M., 2007. {The
  Discovery of 1000 km s$^{-1}$ Outflows in Massive Poststarburst Galaxies at
  $z=0.6$}. \apjl 663, L77--L80.

\bibitem[{{Veilleux} et~al.(2005){Veilleux}, {Cecil}, and
  {Bland-Hawthorn}}]{Veilleux:2005aa}
{Veilleux}, S., {Cecil}, G., {Bland-Hawthorn}, J., 2005. {Galactic Winds}.
  \araa 43, 769--826.

\bibitem[{{Weingartner} and {Draine}(2001)}]{Weingartner:2001aa}
{Weingartner}, J.~C., {Draine}, B.~T., 2001. {Dust Grain-Size Distributions and
  Extinction in the Milky Way, Large Magellanic Cloud, and Small Magellanic
  Cloud}. \apj 548, 296--309.

\bibitem[{{Yan} et~al.(2004){Yan}, {Lazarian}, and {Draine}}]{Yan:2004aa}
{Yan}, H., {Lazarian}, A., {Draine}, B.~T., 2004. {Dust Dynamics in
  Compressible Magnetohydrodynamic Turbulence}. \apj 616, 895--911.

\bibitem[{{York} and {et al.}(2000)}]{York:2000aa}
{York}, D.~G., {et al.}, 2000. {The Sloan Digital Sky Survey: Technical
  Summary}. \aj 120, 1579--1587.

\bibitem[{{York} et~al.(2006){York}, {Khare}, {Vanden Berk}, {Kulkarni},
  {Crotts}, {Lauroesch}, {Richards}, {Schneider}, {Welty}, {Alsayyad}, {Kumar},
  {Lundgren}, {Shanidze}, {Smith}, {Vanlandingham}, {Baugher}, {Hall},
  {Jenkins}, {Menard}, {Rao}, {Tumlinson}, {Turnshek}, {Yip}, and
  {Brinkmann}}]{York:2006aa}
{York}, D.~G., {Khare}, P., {Vanden Berk}, D., {Kulkarni}, V.~P., {Crotts},
  A.~P.~S., {Lauroesch}, J.~T., {Richards}, G.~T., {Schneider}, D.~P., {Welty},
  D.~E., {Alsayyad}, Y., {Kumar}, A., {Lundgren}, B., {Shanidze}, N., {Smith},
  T., {Vanlandingham}, J., {Baugher}, B., {Hall}, P.~B., {Jenkins}, E.~B.,
  {Menard}, B., {Rao}, S., {Tumlinson}, J., {Turnshek}, D., {Yip}, C.-W.,
  {Brinkmann}, J., 2006. {Average extinction curves and relative abundances for
  quasi-stellar object absorption-line systems at $1<=z_{abs}<2$}. \mnras 367,
  945--978.

\bibitem[{{Zibetti} et~al.(2007){Zibetti}, {M{\'e}nard}, {Nestor}, {Quider},
  {Rao}, and {Turnshek}}]{Zibetti:2007aa}
{Zibetti}, S., {M{\'e}nard}, B., {Nestor}, D.~B., {Quider}, A.~M., {Rao},
  S.~M., {Turnshek}, D.~A., 2007. {Optical Properties and Spatial Distribution
  of Mg II Absorbers from SDSS Image Stacking}. \apj 658, 161--184.

\bibitem[{{Zu} et~al.(2011){Zu}, {Weinberg}, {Dav{\'e}}, {Fardal}, {Katz},
  {Kere{\v s}}, and {Oppenheimer}}]{Zu:2011aa}
{Zu}, Y., {Weinberg}, D.~H., {Dav{\'e}}, R., {Fardal}, M., {Katz}, N., {Kere{\v
  s}}, D., {Oppenheimer}, B.~D., 2011. {Intergalactic dust extinction in
  hydrodynamic cosmological simulations}. \mnras 412, 1059--1069.

\bibitem[{{Zubko} et~al.(2004){Zubko}, {Dwek}, and {Arendt}}]{Zubko:2004aa}
{Zubko}, V., {Dwek}, E., {Arendt}, R.~G., 2004. {Interstellar Dust Models
  Consistent with Extinction, Emission, and Abundance Constraints}. \apjs 152,
  211--249.

\bibitem[{{Zubko} et~al.(1996){Zubko}, {Mennella}, {Colangeli}, and
  {Bussoletti}}]{Zubko:1996aa}
{Zubko}, V.~G., {Mennella}, V., {Colangeli}, L., {Bussoletti}, E., 1996.
  {Optical constants of cosmic carbon analogue grains - I. Simulation of
  clustering by a modified continuous distribution of ellipsoids}. \mnras 282,
  1321--1329.

\end{thebibliography}

\end{document}